\documentclass[preprint,review,12pt]{elsarticle}
\usepackage{amssymb}
\usepackage{amsmath}
\usepackage{subcaption}
\usepackage{amsthm}
\usepackage{color}
\usepackage{bm}
\usepackage[linesnumbered, ruled, vlined]{algorithm2e}
\usepackage{multirow}
\usepackage[table]{xcolor}

\journal{arXiv}

\begin{document}
\fontsize{12}{13.5}\selectfont
\captionsetup[figure]{name={Fig.}}

\begin{frontmatter}

\title{Towards a Higher Roofline for Matrix-Vector Multiplication in Matrix-Free HOSFEM} 

\author[1,2]{Zijian Cao}
\author[2]{Qiao Sun}
\author[2]{Tiangong Zhang}
\author[2]{Huiyuan Li\corref{cor1}}

\ead{huiyuan@iscas.ac.cn}
\ead[URL]{https://people.ucas.ac.cn/~lhy}

\cortext[cor1]{Corresponding author}

\affiliation[1]{organization={School of Computer Science and Technology, University of Chinese Academy of Sciences},
            state={Beijing},
            postcode={101408},
            country={China}}

\affiliation[2]{organization={Institute of Software, Chinese Academy of Sciences},
            state={Beijing},
            postcode={100190},
            country={China}}

\begin{abstract}
  Modern GPGPUs provide massive arithmetic throughput, yet many scientific kernels remain limited by memory bandwidth. In particular, repeatedly loading precomputed auxiliary data wastes abundant compute resources while stressing the memory hierarchy. A promising strategy is to replace memory traffic with inexpensive recomputation, thereby alleviating bandwidth pressure and enabling applications to better exploit heterogeneous compute units. Guided by this strategy, we optimize the high-order/spectral finite element method (HOSFEM), a widely used approach for solving PDEs. Its performance is largely determined by AxLocal, a matrix-free kernel for element-local matrix-vector multiplications. In AxLocal, geometric factors dominate memory accesses while contributing minimally to computation, creating a bandwidth bottleneck that caps the performance roofline. To address this challenge, we propose the first practical, low-overhead on-the-fly recomputation of geometric factors for trilinear and parallelepiped elements. This reformulation reduces data movement and raises the achievable roofline, revealing untapped optimization potential for tensor contractions. With hardware-aware techniques including loop unrolling, Tensor Core acceleration, and constant memory utilization, the optimized kernels reach 85\%-100\% of the roofline efficiency. Compared with state-of-the-art implementations in the Nek series, they deliver speedups of 1.74x-4.10x on NVIDIA A100 and 1.99x-3.78x on Hygon K100, leading to a 1.12x-1.40x improvement in the full HOSFEM benchmark. These results demonstrate that combining algorithmic reformulation with hardware-specific tuning can remove long-standing bottlenecks and fully exploit the performance potential of large-scale high-order simulations.
\end{abstract}

\begin{keyword}
matrix-vector multiplication \sep matrix-free kernel \sep tensor-product \sep trilinear element \sep geometric factor \sep GPGPU
\end{keyword}

\end{frontmatter}

\thispagestyle{empty}
\newpage
\setcounter{page}{1}

\section{Introduction}
HOSFEM is widely applied in structural mechanics, computational fluid dynamics (CFD), and electromagnetics for solving partial differential equations (PDEs). Compared with low-order methods, it can achieve a target accuracy with fewer degrees of freedom and lower overall computational cost. Its high compute-to-memory-access ratio, block-structured computation, and localized communication patterns make it highly suitable for Exascale high-performance computing (HPC) architectures. As the core discretization method within the Center for Efficient Exascale Discretizations (CEED) \cite{CEED_website} under the Exascale Computing Project (ECP) \cite{ECP_website}, HOSFEM has been integrated into flagship software frameworks such as MFEM \cite{MFEM_website}, the Nek series \cite{NEK_website}, and libParanumal \cite{libParanumal_website}. Similar to most discretization methods, the linear system from HOSFEM is typically solved using iterative methods (e.g., CG, GMRES). In each iteration, the core operation is a matrix-vector multiplication $\mathbf{Y} = \mathbf{A} \mathbf{X}$, which dominates the overall computational workload, to update the approximate solution until convergence, where $\mathbf{A}$ is the coefficient matrix from the discretization. Apart from the matrix-vector multiplication, the remaining computations mainly consist of vector-level operations.

The key distinction is that HOSFEM partitions the global domain into multiple elements and, rather than explicitly forming and storing the entries of the global matrix $\mathbf{A}$, applies it in an element-wise manner through local products $\mathbf{Y}^{(e)} = \mathbf{A}^{(e)} \mathbf{X}^{(e)}$, referred to in this work as AxLocal. A gather-scatter operation is then used to handle communication between neighboring elements and to assemble the global product $\mathbf{Y}$. The AxLocal operator has been consistently identified as one of the most computationally intensive components in HOSFEM~\cite{Parallel_Communication_Nekbone, Nekbone_GPUs_OpenACC_CUDA, Acceleration_tensor_product_operations_in_FEM, NekRS_first_paper, Optimization_Ax_Nekbone_SC_20_Poster, Accelerate_Nekbone_on_FPGA, Accelerate_axhelm_in_Nekbone_on_A64FX, A64FX_performance, mixed_precision_Nekbone} and serves as a common operator across various applications~\cite{NekRS_first_paper, Neko_first_paper, Large_Scale_turbulence_simulation, wind_energy_simulations, Full_Core_Reactor_Simulations_on_Summit, Exascale_Nuclear_Reactor_Simulations, Nek5000_RS_Performance_on_Advanced_GPU}. As the fifth Back-off Kernel (BK5)~\cite{CEED_BPS_website, HP_SEM_FPGA}, it plays a central role in the CEED co-design effort~\cite{CEED_website, Acceleration_tensor_product_operations_in_FEM}. Analogous to the global-level formulation, AxLocal does not explicitly store the element-level matrices $\mathbf{A}^{(e)}$. Instead, it exploits their tensor-product structure and employs sum-factorization techniques to perform tensor contractions on the input vectors $\mathbf{X}^{(e)}$, followed by applying the element-specific geometric factors. This approach greatly reduces both computational complexity and storage cost. Matrix-free formulations at both the global and element levels form a core of HOSFEM.

Similar to how the computational workload is concentrated in $\mathbf{Y} = \mathbf{A} \mathbf{X}$ and subsequently in AxLocal, the majority of AxLocal’s computational cost lies in tensor contractions. This part has undergone extensive optimization, yielding remarkable results, with AxLocal achieving over 90\% of its theoretical performance limit across various platforms. In contrast, the other component of AxLocal, which involves minimal computational workload and involves the application of element-specific geometric information, has received comparatively little attention. Typically, the geometric data is computed during the setup phase and repeatedly loaded from memory in each iteration. These fixed geometric data account for more than half of the memory accesses in AxLocal, imposing a substantial bandwidth demand at runtime.

Our research is motivated by three key observations: (1) The performance of PDE solvers is often limited by memory bandwidth, making it essential to leverage the surplus peak computing capability of the target platform to alleviate this bottleneck. (2) The trilinear type is the default for hexahedral meshes in real-world HOSFEM applications, since the majority of non-boundary elements fall into this category. (3) The geometric factors of a trilinear element can be recomputed on-the-fly at significantly lower cost. These lead to central questions: What is the performance upper bound of an HOSFEM-based solver, and how can it be achieved for meshes dominated by trilinear elements? In this work, unlike prior studies focusing solely on the Poisson equation for scalar fields, we conduct a comprehensive study covering all subproblems of the Navier-Stokes (NS) equations. Our contributions are as follows:
\begin{itemize}
  \setlength{\itemsep}{0pt}
  \setlength{\parsep}{0pt}
  \setlength{\parskip}{0pt}
  \item We propose a practical strategy to replace memory accesses for geometric factors with low-cost on-the-fly recomputation in HOSFEM, thereby overcoming the bandwidth bottleneck of AxLocal. This represents the first solution with sufficiently low overhead to make recomputation practical.
  \item We develop a time-based roofline performance model to ensure that the effective performance metric excludes the extra computation incurred when memory accesses are replaced by on-the-fly recomputation.
  \item A higher roofline suggests substantial room for further tensor contraction optimization. With optimizations such as Tensor Core acceleration and constant memory utilization, the performance approaches the higher theoretical limit.
  \item A speedup of 1.74x to 4.10x for AxLocal and 1.12x to 1.40x for the full HOSFEM benchmark is achieved for optimized performance on the NVIDIA A100 and Hygon K100.
\end{itemize}

Section~\ref{Background_section} provides the necessary background. Section~\ref{The_Proposed_Method_section} analyzes the performance bottleneck and presents the recomputation algorithms along with the performance model. Implementation and optimization details follow in Section~\ref{Implementation_and_optimization_section}, while Section~\ref{Experiments_section} presents the experimental results. Related work is reviewed in Section~\ref{Related_Work_section}, and the paper concludes with future research directions in Section~\ref{Conclusion_section}.

\section{Background}
\label{Background_section}
\subsection{HOSFEM and Matrix-Free Method}
\label{SE_Discretization_Section}
\begin{figure}[htbp]
  \centering
  \includegraphics[width=1.0\linewidth]{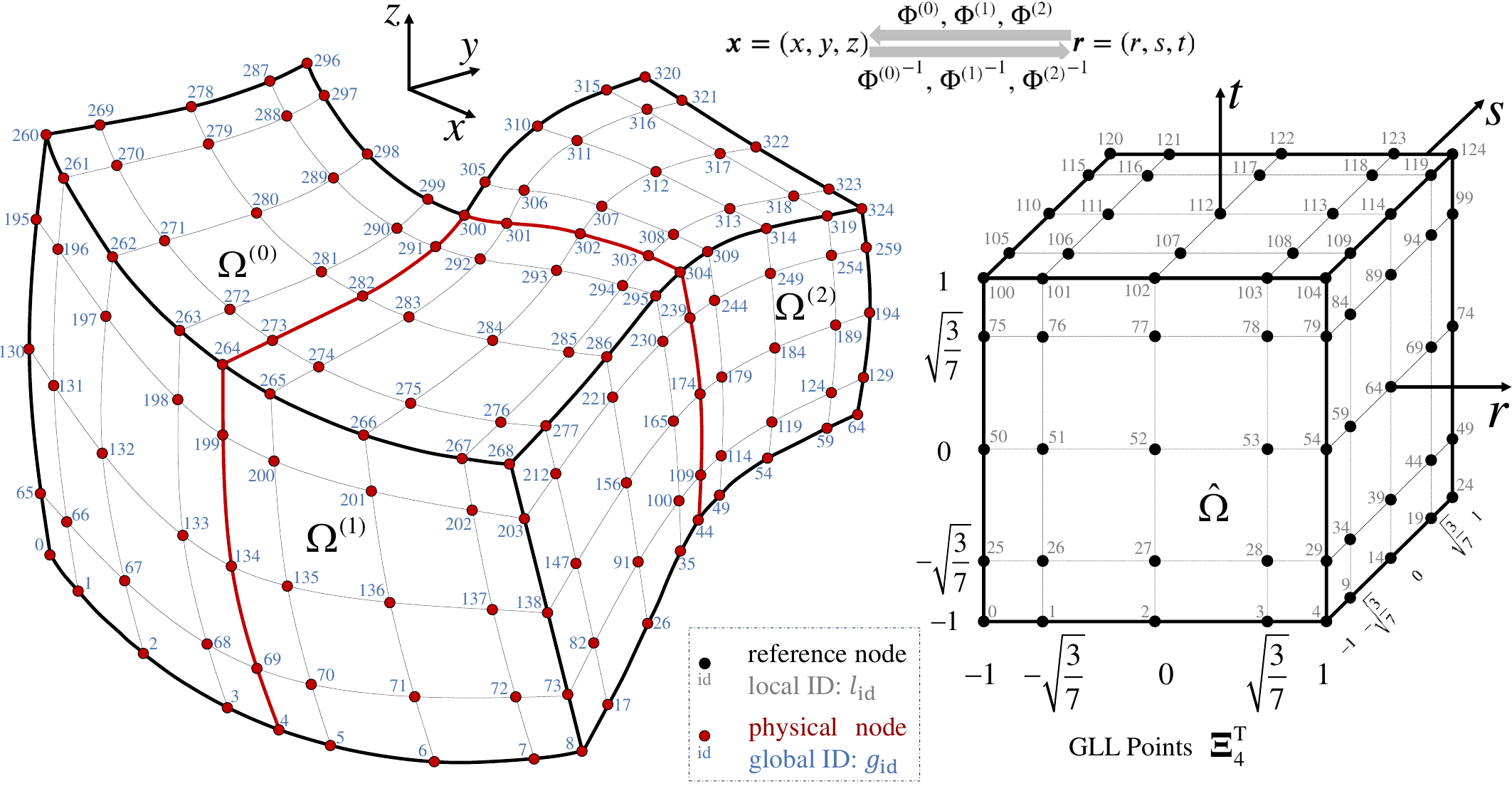}
  \caption{HOSFEM discretization example: $(E, N, \mathcal{N})$=(3,4,325).}
  \label{reference_nodes_to_physical_nodes_fig}
\end{figure}

\begin{table}[h]
  \centering
  \caption{Key concepts in HOSFEM ($0 \leq i, j, k \leq N$).}
  \centerline{\scriptsize
    {
      \begin{tabular}{|c|c|c|}
        \hline
        \textbf{Concept}                              & \textbf{Symbol}              & \textbf{Definition}                                                      \\
        \hline
        Legendre Polynomial                                & $L_{N}(x)$                   & {$\frac{1}{2^N N!} \cdot \frac{\mathrm{d}^N}{{\mathrm{d}x}^N}(x^2 - 1)^N$} \\
        GLL Points: $(\xi_0, \dots, \xi_N)^\top$ & $\mathbf{\Xi}_{N}$           & {zeros of $(1 - x^2)L_{N}'(x)$}                                            \\
        GLL Weights: $(w_0, \dots, w_N)^\top$    & $\mathbf{W}_{N}$             & {$w_i = 2/[N(N + 1)(L_{N}(\xi_i))^2]$}                 \\
        Lagrange Interp. Poly.                        & $\pi_i(x)$                   & {{$\frac{-1}{N(N+1)} \cdot \frac{(1 - x^2)L'_N(x)}{(x - \xi_i)L_N(\xi_i)}$}} \\
        Differentiation Matrix                        & $\hat{\mathbf{D}}_{N}$       & {$\hat{\mathbf{D}}_{N}[i][j] = \pi_j'(\xi_i)$}                             \\
        Identity Matrix                        & $\mathbf{I}$                & size of $(N+1) \times (N+1)$                             \\
        Tensor (Kronecker) Product                        & $\otimes$                  & see Eq.~(4) in \cite{Cite_for_Tensor_product_define}                    \\
        Discrete Gradient $\nabla = {{\begin{bmatrix}
      \partial_r \\
      \partial_s \\
      \partial_t
    \end{bmatrix}}}$ & $\mathbf{D}$ & {{$    \begin{bmatrix}
      \mathbf{D}_r \\
      \mathbf{D}_s \\
      \mathbf{D}_t
    \end{bmatrix} :=
    \begin{bmatrix}
      \mathbf{I} \otimes \mathbf{I} \otimes \hat{\mathbf{D}}_N \\
      \mathbf{I} \otimes \hat{\mathbf{D}}_N \otimes \mathbf{I} \\
      \hat{\mathbf{D}}_N \otimes \mathbf{I} \otimes \mathbf{I}
    \end{bmatrix}$}} \\
        \hline
      \end{tabular}}}
  \label{Key_Concepts_SEM_Table}
\end{table}
Table~\ref{Key_Concepts_SEM_Table} summarizes the key concepts in HOSFEM \cite{High_Order_Methods_Incompressible_Fluid_Flow}. For this paper, it is sufficient to note that once the order $N$ is determined, the vectors, matrices, and polynomial coefficients become fixed constants. For instance, when $N = 2$, $L_2(x) = \frac{3x^2 - 1}{2}$, $\mathbf{\Xi}_{2} = (-1, 0, 1)^\top$, $\mathbf{W}_{2} = (\frac{1}{3}, \frac{4}{3}, \frac{1}{3})^\top$, and $\hat{\mathbf{D}}_{2} = $ {\tiny{$\begin{bmatrix} -1.5 & 2 & -0.5 \\ -0.5 & 0 & 0.5 \\ 0.5 & -2 & 1.5 \end{bmatrix}$}}. Let $N_1 = N + 1$. Then $\mathbf{D}$ is of size $3N_1^3 \times N_1^3$. As shown in Fig.~\ref{reference_nodes_to_physical_nodes_fig}, an example of HOSFEM discretization in $\mathbb{R}^3$ includes:

(1) \textbf{Mesh}: The global domain $\Omega$ is partitioned into $E$ elements, $\Omega^{(e)}$ with $0 \leq e < E$. The reference element $\hat{\Omega} = [-1, 1]^3$ is then mapped onto $\Omega^{(e)}$ through the mapping $\Phi^{(e)}: \hat{\Omega} \to \Omega^{(e)}$.

(2) \textbf{High-Order Nodal Basis}: The reference nodes $\{(\xi_i, \xi_j, \xi_k) \mid 0 \leq i, j, k \leq N\}$ in the reference element are mapped onto the physical nodes of element $e$ through $\Phi^{(e)}$. Each element contains $N_1^3$ nodes, where boundary nodes are shared among neighboring elements. The global domain contains $\mathcal{N}$ unique nodes, satisfying $\mathcal{N} < EN_1^3$. The global-to-local matrix $\mathbf{Q}$, of size $EN_1^3 \times \mathcal{N}$, is a sparse binary matrix defined by $\mathbf{Q}[eN_1^3 + l_\text{id}][g_\text{id}] = 1$ if and only if the $l_\text{id}$-th node of element $e$ has the global ID $g_\text{id}$. Within each element, the geometric deformation at each node introduces seven geometric factors (Section~\ref{Computation_of_Geometric_Factors_subsection}). Once the mesh and nodes are determined, they remain fixed.

After HOSFEM discretization, the PDE defined over $\Omega$ gives rise to a linear system with a coefficient matrix expressed as
\begin{equation}
  \mathbf{A} = \mathbf{Q}^\top
  \begin{bmatrix}
    \mathbf{A}^{(0)} &                  &        &                      \\
                     & \mathbf{A}^{(1)} &        &                      \\
                     &                  & \ddots &                      \\
                     &                  &        & \mathbf{A}^{(E - 1)}
  \end{bmatrix}
  \mathbf{Q} \ \ \in \ \ \mathbb{R}^{\mathcal{N} \times \mathcal{N}},
\end{equation}
where each element-local matrix $\mathbf{A}^{(e)}$ has dimension $N_1^3 \times N_1^3$.

\begin{figure}[htbp]
  \centering
  \includegraphics[width=1.0\linewidth]{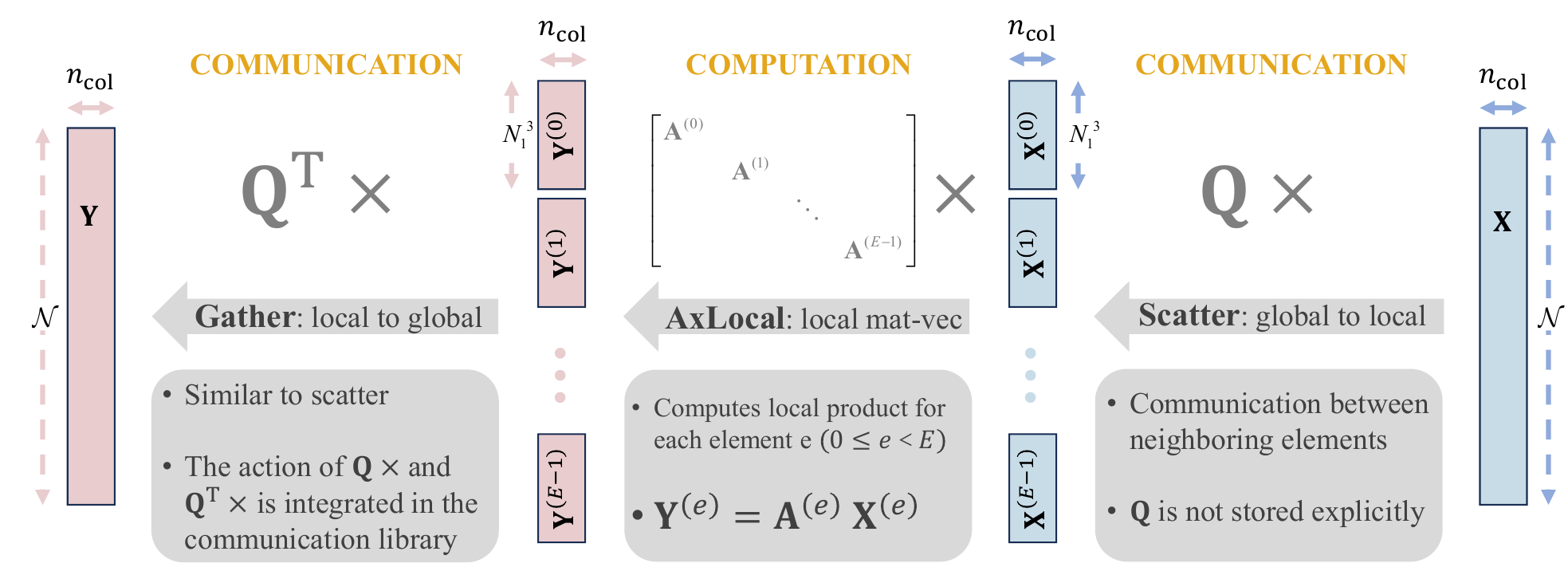}
  \caption{Matrix-free method for computing $\mathbf{Y} = \mathbf{A} \mathbf{X}$ via gather-scatter and AxLocal.}
  \label{matrix_free_kernels_fig}
\end{figure}

To compute $\mathbf{Y} = \mathbf{A} \mathbf{X}$ in each iteration, matrix-based methods typically store $\mathbf{A}$ in a sparse format and perform sparse matrix-vector multiplication (SpMV). Note that $\mathbf{X}$ and $\mathbf{Y}$ contain either one or three columns, i.e., $n_\mathrm{col} \in \{1,3\}$, depending on whether the PDE solution is a scalar or vector field. In contrast, as illustrated in Fig.~\ref{matrix_free_kernels_fig}, HOSFEM does not store $\mathbf{A}$, nor $\mathbf{A}^{(e)}$ or $\mathbf{Q}$. Instead, the computation proceeds through a sequence of matrix-free operations, where the gather-scatter and AxLocal kernels are decoupled, handling communication and computation, respectively. Apart from the preprocessing stage, these two kernels constitute the most critical matrix-free components in the iterative process, since all other operations reduce to vector operations. The analysis now turns to AxLocal, the core computational kernel in matrix-free HOSFEM.

\subsection{AxLocal: Core Computation in HOSFEM}
The Helmholtz equation is defined as
\begin{equation}
  \label{Helmholtz_equation}
  - \nabla \cdot (\lambda_0(\bm{x}) \nabla \bm{u}(\bm{x})) + \lambda_1(\bm{x}) \bm{u}(\bm{x}) = \bm{f}(\bm{x}), \quad\bm{x} \in \Omega \subset \mathbb{R}^3,
\end{equation}
where $\bm{u}(\bm{x})$ is the unknown scalar or vector field, $\bm{f}(\bm{x})$ is the source term, and $\lambda_0(\bm{x})$, $\lambda_1(\bm{x})$ are given scalar coefficient fields.

For instance, considering the Helmholtz equation (or the Poisson equation with $\lambda_0(\bm{x}) \equiv 1$ and $\lambda_1(\bm{x}) \equiv 0$), $\mathbf{A}^{(e)}$ is given by
{\small
\begin{equation}
  \label{local_matrix_structure}
  \mathbf{A}^{(e)} = \mathbf{D}^\top
  \begin{bmatrix}
      \mathbf{\Lambda}_0^{(e)} &                      &                      \\
                           & \mathbf{\Lambda}_0^{(e)} &                      \\
                           &                      & \mathbf{\Lambda}_0^{(e)}
    \end{bmatrix}
    \begin{bmatrix}
      \mathbf{G}_{00}^{(e)} & \mathbf{G}_{01}^{(e)} & \mathbf{G}_{02}^{(e)} \\
      \mathbf{G}_{01}^{(e)} & \mathbf{G}_{11}^{(e)} & \mathbf{G}_{12}^{(e)} \\
      \mathbf{G}_{02}^{(e)} & \mathbf{G}_{12}^{(e)} & \mathbf{G}_{22}^{(e)}
    \end{bmatrix}
  \mathbf{D} + \mathbf{\Lambda}_1^{(e)} \mathbf{G}_\mathrm{wj}^{(e)},
\end{equation}
}
where each term is defined for element $e$ as follows: (1) $\mathbf{\Lambda}_0^{(e)}$ and $\mathbf{\Lambda}_1^{(e)}$ are diagonal matrices of size $N_1^3 \times N_1^3$, with diagonal entries given by the nodal values of $\lambda_0(\bm{x})$ and $\lambda_1(\bm{x})$, respectively; (2) the geometric factors of element $e$ form seven diagonal matrices of size $N_1^3 \times N_1^3$: $\mathbf{G}_{00}^{(e)}$, $\mathbf{G}_{01}^{(e)}$, $\mathbf{G}_{02}^{(e)}$, $\mathbf{G}_{11}^{(e)}$, $\mathbf{G}_{12}^{(e)}$, $\mathbf{G}_{22}^{(e)}$, and $\mathbf{G}_\mathrm{wj}^{(e)}$; (3) in the Poisson case, $\mathbf{\Lambda}_0^{(e)}$, $\mathbf{\Lambda}_1^{(e)}$, and $\mathbf{G}_\mathrm{wj}^{(e)}$ do not appear.

Algorithm~\ref{AxLocal_algorithm} illustrates the computation of $\mathbf{Y}^{(e)} = \mathbf{A}^{(e)} \mathbf{X}^{(e)}$ without explicitly constructing $\mathbf{A}^{(e)}$. This is the standard matrix-free AxLocal kernel framework, where each component of $\mathbf{A}^{(e)}$ acts directly on the input vector(s). Note that the notation $[i,j,k]$ denotes the $(i + jN_1 + kN_1^2)$-th entry of an $N_1^3$-dimensional vector or diagonal matrix. The primary computational workload (Lines 7-8, 14-15) arises from six tensor contractions, each corresponding to the multiplication of a tensor-product (e.g., $\mathbf{D}_r$, defined in Table~\ref{Key_Concepts_SEM_Table}) with a vector (e.g., $\mathbf{x}$), which requires $2N_1^4$ floating-point operations. The tensor contraction is derived from the sum factorization technique~\cite{High_Order_Methods_Incompressible_Fluid_Flow, NekRS_first_paper, wind_energy_simulations}, which exploits the local tensor-product structure to reduce the computational complexity of $\mathbf{Y}^{(e)} = \mathbf{A}^{(e)} \mathbf{X}^{(e)}$ from $\mathcal{O}(N_1^6)$ to $\mathcal{O}(N_1^4)$. This reduction is the fundamental source of high performance in HOSFEM.

\begin{algorithm}
  \fontsize{9}{9}\selectfont
  \caption{Matrix-free AxLocal kernel: $\mathbf{Y}^{(e)} = \mathbf{A}^{(e)} \mathbf{X}^{(e)}$}
  \label{AxLocal_algorithm}
  \SetKwInOut{Input}{\textbf{Input}}
  \SetKwInOut{Output}{\textbf{Output}}
  \SetKwInOut{Init}{\textbf{Init}}
  \SetKw{KwDStep}{\textbf{step}}
  \SetKw{KwDAnd}{\textbf{and}}
  \SetKw{KwDIn}{\textbf{in}}
  \Input{$\mathbf{\Lambda}_0^{(e)}, \mathbf{\Lambda}_1^{(e)}, \mathbf{G}_{00}^{(e)}, ..., \mathbf{G}_{22}^{(e)}, \mathbf{G}_\mathrm{wj}^{(e)}, \mathbf{X}^{(e)}, \hat{\mathbf{D}}_N$.}
  \Output{$\mathbf{Y}^{(e)}$.}
  \For{$0 \leq e < E$}{
    \For{$\mathrm{each\ column\ pair\ } (\mathbf{x}, \mathbf{y})\mathrm{\ of\ }\mathbf{X}^{(e)}\mathrm{\ and\ }\mathbf{Y}^{(e)}$}{
      \For{$0 \leq \mathrm{i, j, k} \leq N$}{
        $g_{00} = \mathbf{G}_{00}^{(e)}$[i,j,k]; $g_{01} = \mathbf{G}_{01}^{(e)}$[i,j,k]; $g_{02} = \mathbf{G}_{02}^{(e)}$[i,j,k]\;
        $g_{11} = \mathbf{G}_{11}^{(e)}$[i,j,k]; $g_{12} = \mathbf{G}_{12}^{(e)}$[i,j,k]; $g_{22} = \mathbf{G}_{22}^{(e)}$[i,j,k]\;
        $\lambda_0 = \mathbf{\Lambda}_0^{(e)}$[i,j,k]\;
        $x_{0} = \sum_\mathrm{n = 0}^{N} \hat{\mathbf{D}}_N$[i][n]$\mathbf{x}$[n,j,k]; $x_{1} = \sum_\mathrm{n = 0}^{N} \hat{\mathbf{D}}_N$[j][n]$\mathbf{x}$[i,n,k]\;
        $x_{2} = \sum_\mathrm{n = 0}^{N} \hat{\mathbf{D}}_N$[k][n]$\mathbf{x}$[i,j,n]\tcp*[r]{\color{gray}$\mathbf{D}_r\mathbf{x}\ \mathbf{D}_s\mathbf{x}\ \mathbf{D}_t\mathbf{x}$}
        \tcp{\color{gray}Compute intermediate results: $\mathbf{r}$, $\mathbf{s}$, $\mathbf{t}$}
        $\mathbf{r}$[i,j,k] $= \lambda_0(g_{00}x_{0} + g_{01}x_{1} + g_{02}x_{2})$\;
        $\mathbf{s}$[i,j,k] $= \lambda_0(g_{01}x_{0} + g_{11}x_{1} + g_{12}x_{2})$\;
        $\mathbf{t}$[i,j,k] $= \lambda_0(g_{02}x_{0} + g_{12}x_{1} + g_{22}x_{2})$\;
      }
      \For{$0 \leq \mathrm{i, j, k} \leq N$}{
        $g_\mathrm{wj} = \mathbf{G}_\mathrm{wj}^{(e)}$[i,j,k]; $\lambda_1 = \mathbf{\Lambda}_1^{(e)}$[i,j,k]\;
        $y_{0} = \sum_\mathrm{n = 0}^{N} \hat{\mathbf{D}}_N$[n][i]$\mathbf{r}$[n,j,k]; $y_{1} = \sum_\mathrm{n = 0}^{N} \hat{\mathbf{D}}_N$[n][j]$\mathbf{s}$[i,n,k]\;
        $y_{2} = \sum_\mathrm{n = 0}^{N} \hat{\mathbf{D}}_N$[n][k]$\mathbf{t}$[i,j,n]\tcp*[r]{\color{gray}$\mathbf{D}_r^\top\mathbf{r}\ \mathbf{D}_s^\top\mathbf{s}\ \mathbf{D}_t^\top\mathbf{t}$}
        $\mathbf{y}$[i,j,k] $= y_{0} + y_{1} + y_{2} + \lambda_1\cdot g_\mathrm{wj}\cdot \mathbf{x}$[i,j,k]\;
      }
    }
  }
\end{algorithm}

Through tensor contraction, HOSFEM transforms sparse tensor computations into dense matrix multiplications, which makes it particularly well-suited for modern hardware architectures. Nevertheless, AxLocal may still be memory-bound due to accesses to geometric factors, as discussed in Section~\ref{Performance_bottleneck_analysis_of_AxLocal_subsection}. To further analyze the mesh structure and its impact on geometric factor optimization, the following discussion focuses on trilinear elements.

\subsection{Dominance of Trilinear Elements in Practice}
\label{Trilinear_Elements_subsection}
\begin{figure}
  \centering
  \includegraphics[width=0.6\linewidth]{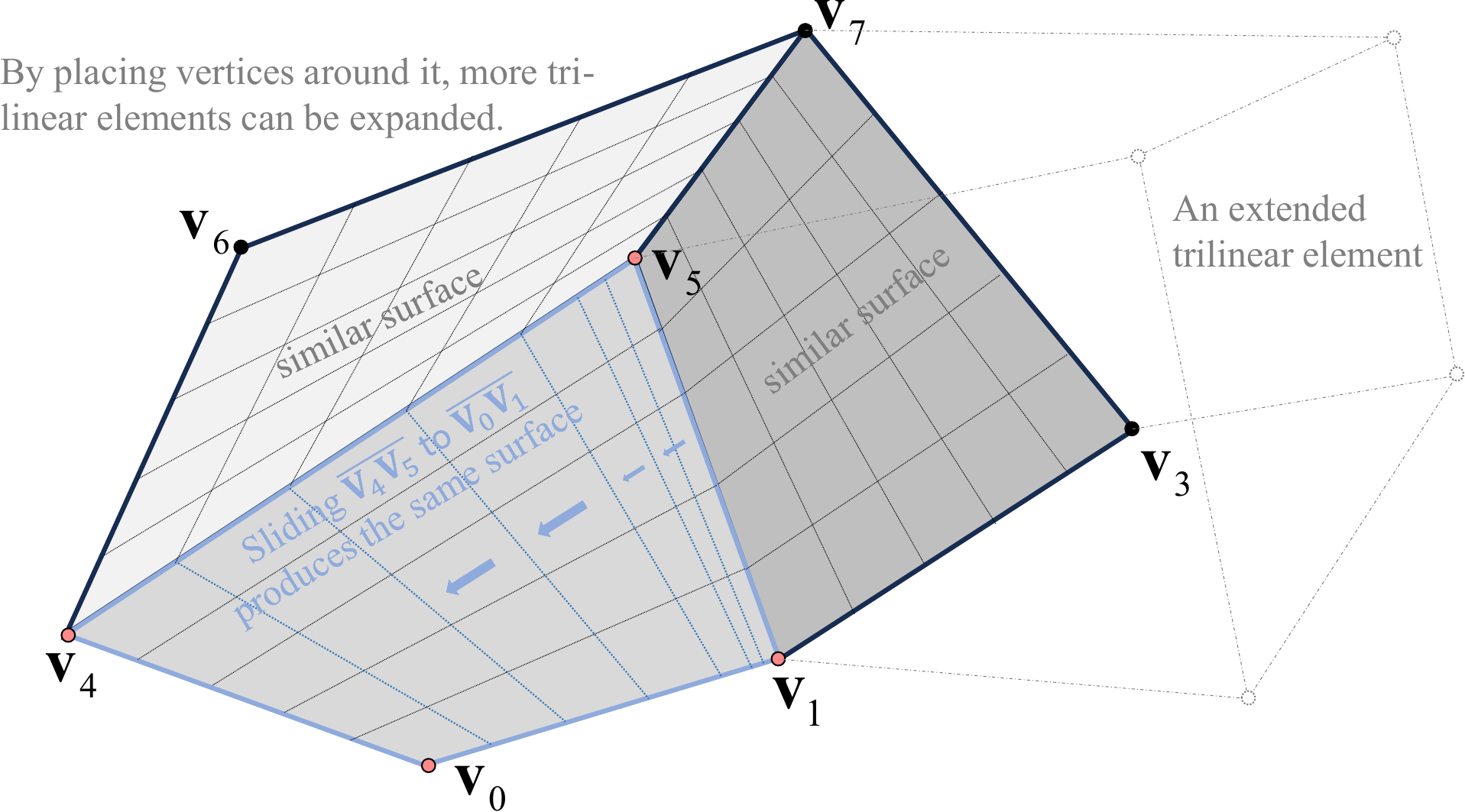}
  \caption{A trilinear element defined by eight vertices, where each edge is a straight line and each face is a ruled surface. For example, sliding $\overline{\mathbf{v}_1\mathbf{v}_5}$ to $\overline{\mathbf{v}_0\mathbf{v}_4}$ generates such a surface.}
  \label{trilinear_element_fig}
\end{figure}

\begin{figure}
  \centering
  \includegraphics[width=0.64\linewidth]{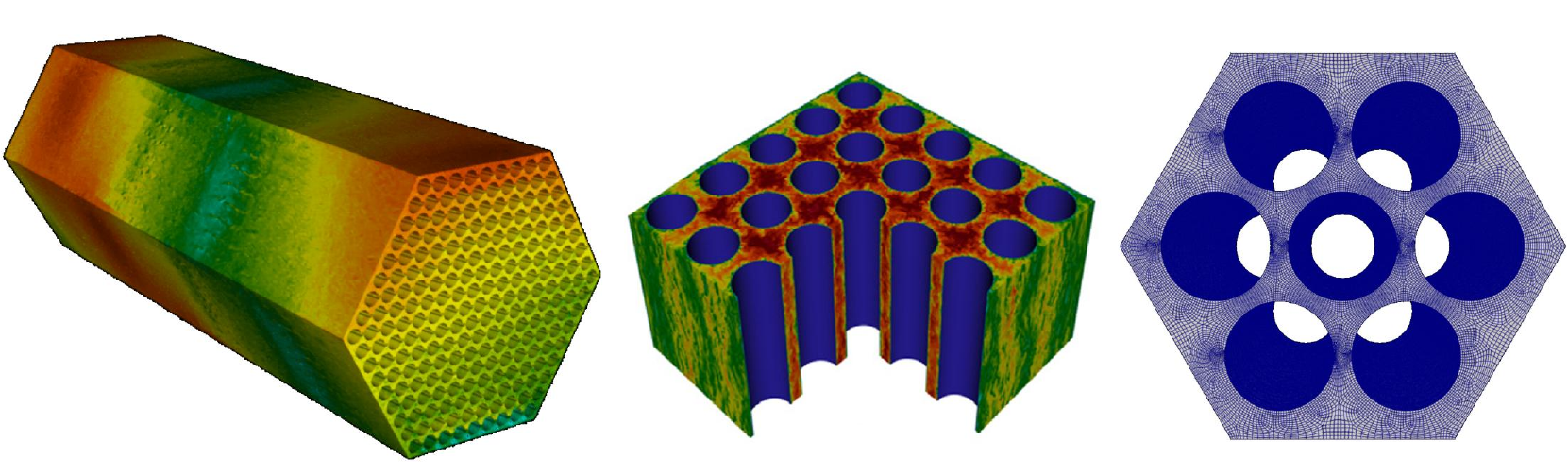}
  \caption{Mesh for a nuclear fuel rod coolant flow simulation.}
  \label{fuel_rod_fig}
\end{figure}

A trilinear element is defined by eight vertex coordinates $\mathbf{v}_i \in \mathbb{R}^3$ $(0 \leq i \leq 7)$ and the mapping $\Phi(r, s, t) := \sum_{i = 0}^{7} \sigma_i(r, s, t)\mathbf{v}_i$. The basis functions $\sigma_i(r, s, t)$ are the standard trilinear shape functions. Complete definitions of all $\sigma_i$ can be found in~\cite{An_efficient_3D_enhanced_strain_element, Trilinear_hexahedral_finite_elements_with_higher_order}. For example,
\[
\sigma_0(r, s, t) = \tfrac{1}{8} (1 - r)(1 - s)(1 - t), \quad \sigma_1(r, s, t) = \tfrac{1}{8} (1 + r)(1 - s)(1 - t).
\]
As a fundamental Q1 element with linear shape functions, the trilinear element has been widely adopted in both geometric modeling and FEM~\cite{A_performance_study_of_tetrahedral_and_hexahedral_elements, Trilinear_hexahedral_finite_elements_with_higher_order, An_efficient_3D_enhanced_strain_element, Higher_order_hierarchical_curved_hexahedral_vector, Improved_versions_of_assumed_enhanced_strain_tri_linear_elements, Poly_Spline_FEM}. As shown in Fig.~\ref{trilinear_element_fig}, a trilinear element is uniquely defined by eight freely chosen vertices~\cite{High_Order_Methods_Incompressible_Fluid_Flow}, from which the edges, faces, and overall geometry are uniquely determined. This flexibility enables trilinear elements to conform to complex geometries and efficiently tessellate the global domain $\Omega$. In practical meshes, the interior of $\Omega$ is often entirely filled with trilinear elements. When higher geometric fidelity is required, higher-order elements such as Q2 or Q3~\cite{Poly_Spline_FEM} are typically employed near the boundary region. Since the surfaces of internal elements do not belong to the physical boundary $\partial\Omega$, their geometric shape does not affect the overall solution accuracy. Mesh statistics (Fig.~\ref{fuel_rod_fig}) from a representative HOSFEM application~\cite{NekRS_first_paper, VisIt_website} show that only 11\% of the elements lie on the boundary, confirming that trilinear elements are sufficiently flexible and form the dominant portion of meshes in real-world applications.


\section{Algorithm Design and Performance Modeling}
\label{The_Proposed_Method_section}
\subsection{Bottleneck Analysis of AxLocal}
\label{Performance_bottleneck_analysis_of_AxLocal_subsection}
Table~\ref{table_workload_memory_access} summarizes the per-element FLOP count and memory access volume for the AxLocal kernels. As shown in Algorithm~\ref{AxLocal_algorithm}, each element requires $12N_1^4$ FLOPs for six tensor contractions, $15N_1^3$ FLOPs for geometric factors (Lines 9-11), and an additional $5N_1^3$ FLOPs in the Helmholtz case ($\lambda_0, \lambda_1, g_\mathrm{wj}$). The total cost is further scaled by the number of field components $n_\mathrm{col}$ (Line 2). Memory access includes input/output vectors, the differentiation matrix, geometric factors, and, in the Helmholtz case, additional coefficient fields, all scaled by the word length (FPSize). Notably, geometric factors account for a substantial portion, which requires $6N_1^3 \cdot \text{FPSize}$ in the Poisson case and $7N_1^3 \cdot \text{FPSize}$ in the Helmholtz case.
\begin{table}[!htb]
  \caption{Per-element computational workload ($\mathtt{F}$) and memory access requirements ($\mathtt{M}$) of the standard AxLocal kernels.}
  \label{table_workload_memory_access}
  \centering
  \scriptsize
  \begin{tabular}{|c|c|c|}
  \hline
  \textbf{Kernel Type} & $\mathtt{F}_\mathrm{ax}$ (\textbf{FLOP}) & $\mathtt{M}_\mathrm{orig}$ (\textbf{Byte}) \\
  \hline
  Poisson, $n_\mathrm{col} = 1$ & $12N_1^4 + 15N_1^3$ & $(8N_1^3 + N_1^2) \cdot \text{FPSize}$ \\
  Helmholtz, $n_\mathrm{col} = 1$ & $12N_1^4 + 20N_1^3$ & $(11N_1^3 + N_1^2) \cdot \text{FPSize}$ \\
  Poisson, $n_\mathrm{col} = 3$ & $36N_1^4 + 45N_1^3$ & $(12N_1^3 + N_1^2) \cdot \text{FPSize}$ \\
  Helmholtz, $n_\mathrm{col} = 3$ & $36N_1^4 + 60N_1^3$ & $(15N_1^3 + N_1^2) \cdot \text{FPSize}$ \\
  \hline
  \end{tabular}
\end{table}

\begin{figure}
  \centering
  \includegraphics[width=0.45\linewidth]{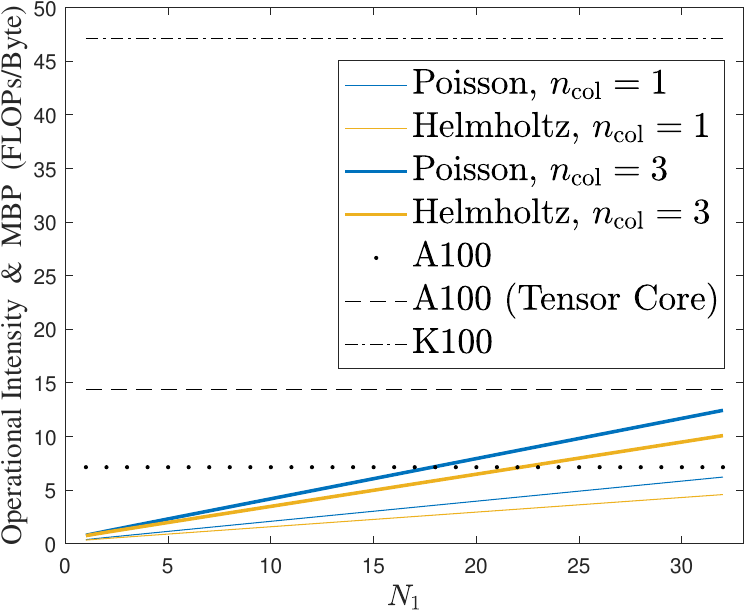}
  \caption{Operational intensity of AxLocal kernels and machine balance points (FP64).}
  \label{operational_intensity_fig}
\end{figure}

The operational intensity is defined as $\mathtt{I} = \mathtt{F}/\mathtt{M}$. The machine balance point (MBP) is given by $\text{MBP} = \mathtt{P}/\mathtt{B}$, where $\mathtt{P}$ denotes the theoretical peak performance and $\mathtt{B}$ the maximum achievable memory bandwidth. MBP thus denotes the minimum operational intensity required to attain peak performance~\cite{Roofline_model_first_paper}. Fig.~\ref{operational_intensity_fig} depicts the operational intensity of four AxLocal kernels and compares them with the MBPs of the target platforms (FPSize = 8).

Fig.~\ref{operational_intensity_fig} shows that the operational intensity of the AxLocal kernel (Poisson, $n_\mathrm{col}=3$) does not exceed an MBP threshold until $N_1=18$. Two key observations follow: (1) the operational intensity grows approximately linearly with $N$; and (2) in practice, AxLocal is predominantly constrained by memory bandwidth. Consequently, under the classical roofline model, where $\mathtt{T}_{\mathrm{cmp}}$ and $\mathtt{T}_{\mathrm{mem}}$ denote computation and memory access time respectively,
\begin{equation}
\label{classical_roofline_performance_model}
  \mathtt{R} = \frac{\mathtt{F}}{\max(\mathtt{T}_{\mathrm{cmp}}, \mathtt{T}_{\mathrm{mem}})} = \frac{\mathtt{F}}{\max\left({\mathtt{F}}/{\mathtt{P}}, {\mathtt{M}}/{\mathtt{B}}\right)} = \min(\mathtt{P}, \mathtt{I} \cdot \mathtt{B}).
\end{equation}
Existing AxLocal implementations~\cite{Acceleration_tensor_product_operations_in_FEM, Optimization_Ax_Nekbone_SC_20_Poster} are already close to their theoretical ceilings, implying that further gains from tensor contraction optimizations alone are unlikely. Achieving performance beyond this limit requires reducing memory traffic. In particular, geometric factors constitute a substantial portion of the memory traffic but contribute only lower-order terms ($\mathcal{O}(N_1^3)$) to the overall FLOP count. Like most prior works, this paper has thus far treated geometric factors as abstract constants. Section~\ref{Computation_of_Geometric_Factors_subsection} explicitly examines how they are computed.

\subsection{Evaluation of Geometric Factors}
\label{Computation_of_Geometric_Factors_subsection}
The geometric factors are related to the Jacobian matrix. For each element $e$, the mapping $\Phi^{(e)}: \hat{\Omega} \to \Omega^{(e)}$, $(r, s, t) \mapsto (x, y, z)$ uniquely determines the Jacobian matrix function:
\begin{equation}
  \label{Jacobian_matrix_function}
  \mathbf{J}^{\Phi^{(e)}}(r, s, t) :=
  \begin{bmatrix}
         \nabla x, \nabla y, \nabla z
       \end{bmatrix}^\top = 
  \begin{bmatrix}
        \partial_r x & \partial_r y & \partial_r z \\
        \partial_s x & \partial_s y & \partial_s z \\
        \partial_t x & \partial_t y & \partial_t z
   \end{bmatrix}^\top.
\end{equation}
This defines a $3 \times 3$ matrix at any point in $\hat{\Omega}$. Accordingly, the Jacobian evaluated at the physical node $(e, i, j, k)$ is given by
\begin{equation}
  \label{Jacobian_matrix_at_node}
  \mathbf{J}_{ijk}^{(e)} := \mathbf{J}^{\Phi^{(e)}}(\xi_i, \xi_j, \xi_k).
\end{equation}
The seven geometric factors at each physical node are defined as:
\begin{equation}
  \label{Jacobian_matrix_to_geometric_factors}
  w_i w_j w_k |\mathbf{J}_{ijk}^{(e)}| \cdot {\mathbf{J}_{ijk}^{(e)}}^{-1} {\mathbf{J}_{ijk}^{(e)}}^\mathrm{-T} \quad\text{and}\quad w_i w_j w_k |\mathbf{J}_{ijk}^{(e)}|,
\end{equation}
where the latter is a scalar and the former is a symmetric $3 \times 3$ matrix involving only six components. The cost of computing geometric factors from the Jacobian matrix, as defined in Eq.~\eqref{Jacobian_matrix_to_geometric_factors}, is well established, requiring a $3 \times 3$ determinant, a matrix inverse, and a matrix multiplication. The objective, therefore, reduces to the efficient computation of the Jacobian matrix itself.

\begin{figure}[htbp]
  \centering
  \includegraphics[width=0.89\linewidth]{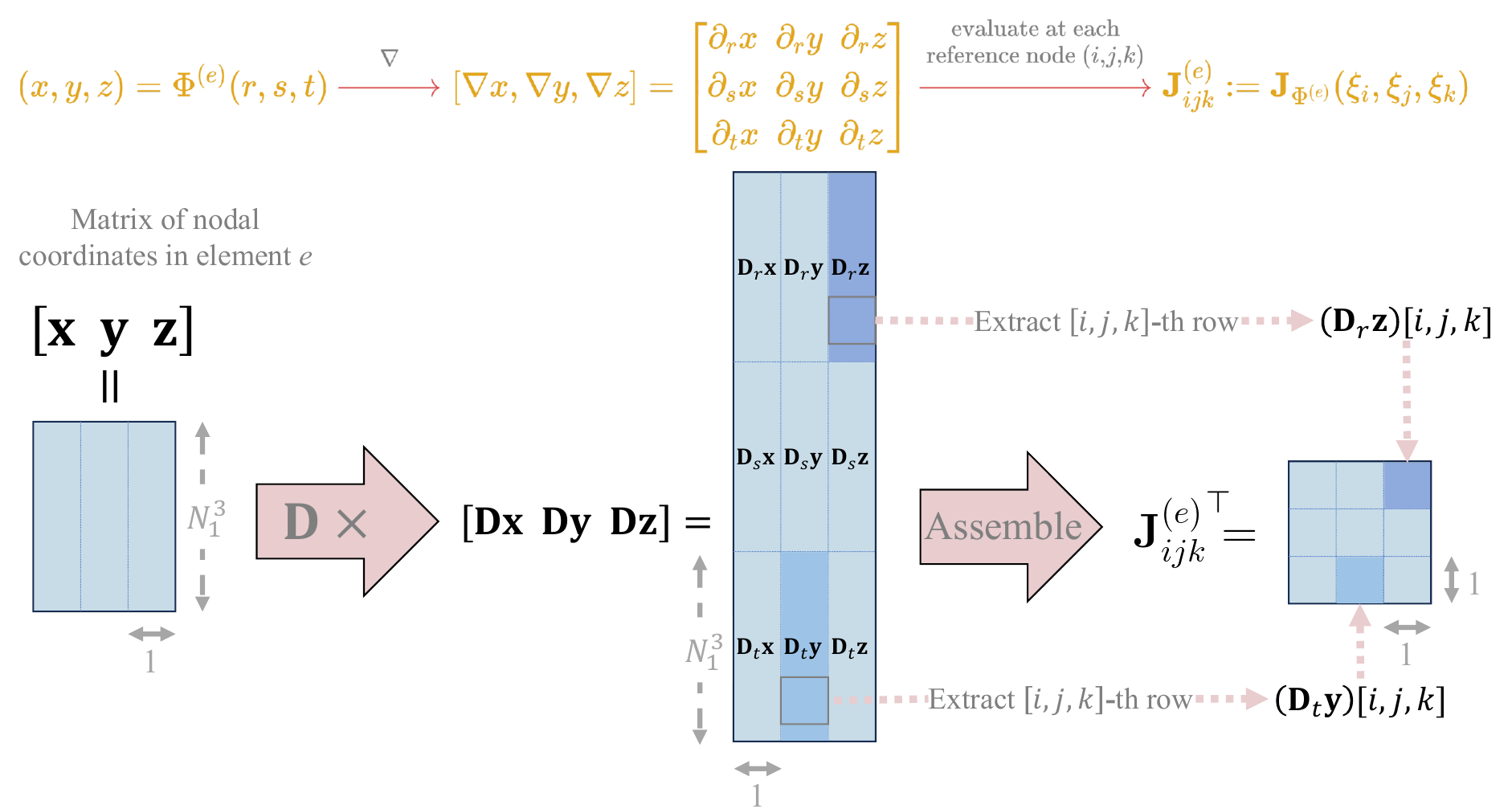}
  \caption{Standard discrete evaluation of the Jacobian matrix in HOSFEM, consistent with its analytical definition.}
  \label{compute_geo_fig}
\end{figure}
However, computing the Jacobian matrix directly from its definition in Equations~\eqref{Jacobian_matrix_function} and \eqref{Jacobian_matrix_at_node} requires analytical expressions of the mapping $\Phi^{(e)}$ and its Jacobian, which are generally unavailable in practice. As illustrated in Fig.~\ref{compute_geo_fig}, the standard discrete procedure instead evaluates the Jacobian numerically from the nodal coordinates, since only the values of $\Phi^{(e)}$ at the nodes are known. The procedure begins by applying the discrete gradient operator $\mathbf{D}$ to the coordinate matrix, yielding nine vector blocks. By extracting the $[i,j,k]$-th entry from each of these vectors, the Jacobian matrix at node $(e, i, j, k)$ is then assembled. This procedure is carried out during the setup stage, immediately after loading the mesh file.

For a single element, computing the Jacobian matrix at all its nodes using this procedure requires $3N_1^3$ memory accesses to the nodal coordinates and nine tensor contractions, totaling $18N_1^4$ FLOPs. Incorporating this overhead, together with $N_1^3$ evaluations of Eq.~\eqref{Jacobian_matrix_to_geometric_factors}, into the AxLocal kernel to substitute merely $6N_1^3$ or $7N_1^3$ memory references for geometric factors is not computationally efficient. However, when $\Phi^{(e)}$ has a simple analytical expression, $\mathbf{J}^{\Phi^{(e)}}$ typically admits a simple form as well. In this case, Eq.~\eqref{Jacobian_matrix_at_node} allows the Jacobian matrix at each node to be computed in $\mathcal{O}(1)$ time.

The analytical approach does not apply to all elements. The approach must therefore target a specific type of hexahedral element that satisfies two conditions: (1) it is sufficiently simple to admit an analytical expression of its Jacobian matrix, and (2) it is sufficiently flexible for broad applicability across meshes. As demonstrated in Section~\ref{Trilinear_Elements_subsection}, the trilinear element satisfies the second requirement, and Section~\ref{trilinear_recalculation_subsection} further shows that it also meets the first, thereby making it an ideal candidate.

\subsection{Low-Cost Recomputation of Geometric Factors for Trilinear Elements}
\label{trilinear_recalculation_subsection}
Let $\mathbf{V}^{(e)} = (\mathbf{v}_0^\top, \dots, \mathbf{v}_7^\top)^\top$ be an $8 \times 3$ matrix containing the coordinates of eight vertices of a trilinear element $e$. Its trilinear mapping can then be written in the compact tensor-product form:
\begin{equation}
  \label{trilinear_mapping_compact_form}
  \Phi^{(e)}(r, s, t) = \frac{1}{8} [(1-t, 1+t) \otimes (1-s, 1+s) \otimes (1-r, 1+r)] \mathbf{V}^{(e)}.
\end{equation}
Substituting Eq.~\eqref{trilinear_mapping_compact_form} into Eq.~\eqref{Jacobian_matrix_function}, the Jacobian matrix function for the trilinear element is obtained as
{\small
\begin{equation}
  \label{trilinear_Jacobian}
  \mathbf{J}^{\Phi^{(e)}}(r, s, t) =
  \frac{1}{8} 
  \left(\begin{bmatrix}
    (1-t, 1+t) \otimes (1-s, 1+s) \otimes (-1, 1) \\ (1-t, 1+t) \otimes (-1, 1) \otimes (1-r, 1+r) \\ (-1, 1) \otimes (1-s, 1+s) \otimes (1-r, 1+r)
  \end{bmatrix}{\mathbf{V}^{(e)}}\right)^\top.
\end{equation}
}

However, direct substitution of $(\xi_i, \xi_j, \xi_k)$ ($0 \leq i, j, k \leq N$) into Eq.~\eqref{trilinear_Jacobian} to compute $\mathbf{J}_{ijk}^{(e)}$ is computationally impractical. Even without considering the cost of expanding the tensor-product, the evaluation at a single node already involves a $3 \times 3 \times 8$ matrix multiplication, incurring a computational cost of 144 FLOPs. To improve efficiency, it is essential to analyze the structure of the expression, identify common subexpressions, and reuse invariant components. Based on these insights, we propose Algorithm~\ref{algorithm_geometric_factors_trilinear}, whose key design principles are as follows:

\begin{algorithm}
  \fontsize{9}{9}\selectfont
  \caption{Geometric factors recomputation (trilinear)}
  \label{algorithm_geometric_factors_trilinear}
  \SetKwInOut{Input}{\textbf{Input}}
  \SetKwInOut{Output}{\textbf{Output}}
  \SetKwInOut{Init}{\textbf{Init}}
  \SetKw{KwDStep}{\textbf{step}}
  \SetKw{KwDAnd}{\textbf{and}}
  \SetKw{KwDIn}{\textbf{in}}
  \SetKw{KwDReturn}{\textbf{return}}
  \SetKw{KwDCall}{\textbf{call}}
  \Input{$\mathbf{V}^{(e)}$: vertex coordinates of a trilinear element e}
  \For{$0 \leq \mathrm{i} \leq N$}{
    r = $\xi_i$;\quad $\mathrm{r}_0$ = 1 - r;\quad $\mathrm{r}_1$ = 1 + r\tcp*[r]{\color{gray}$\mathbf{V}^{(e)}$ is abbreviated as $\mathbf{V}$}
    \For{$0 \leq \mathrm{c} < 3$}{
      tmp1 = $\mathrm{r}_0 (\mathbf{V}$[1][c] - $\mathbf{V}$[0][c]) + $\mathrm{r}_1 (\mathbf{V}$[3][c] - $\mathbf{V}$[2][c])\;
      tmp2 = $\mathrm{r}_0 (\mathbf{V}$[5][c] - $\mathbf{V}$[4][c]) + $\mathrm{r}_1 (\mathbf{V}$[7][c] - $\mathbf{V}$[6][c])\;
      tmp3 = $\mathrm{r}_0 (\mathbf{V}$[2][c] - $\mathbf{V}$[0][c]) + $\mathrm{r}_1 (\mathbf{V}$[3][c] - $\mathbf{V}$[1][c])\;
      tmp4 = $\mathrm{r}_0 (\mathbf{V}$[6][c] - $\mathbf{V}$[4][c]) + $\mathrm{r}_1 (\mathbf{V}$[7][c] - $\mathbf{V}$[5][c])\;
      $\mathbf{E}_0$[i][c] = tmp1 + tmp2; $\mathbf{E}_1$[i][c] = tmp2 - tmp1\;
      $\mathbf{F}_0$[i][c] = tmp3 + tmp4; $\mathbf{F}_1$[i][c] = tmp4 - tmp3\;
    }
  }
  \For{$0 \leq \mathrm{i,j} \leq N$}{
    r = $\xi_i$;\quad $\mathrm{r}_0$ = 1 - r;\quad $\mathrm{r}_1$ = 1 + r;\quad s = $\xi_j$;\quad $\mathrm{s}_0$ = 1 - s;\quad $\mathrm{s}_1$ = 1 + s\;
      \For{$0 \leq \mathrm{c} < 3$}{
        J$_{\mathrm{c}2}$ = $\mathrm{r}_0 \mathrm{s}_0(\mathbf{V}$[4][c] - $\mathbf{V}$[0][c]) + $\mathrm{r}_1 \mathrm{s}_0(\mathbf{V}$[5][c] - $\mathbf{V}$[1][c])\\\ \quad\ + $\mathrm{r}_1 \mathrm{s}_1(\mathbf{V}$[7][c] - $\mathbf{V}$[3][c]) + $\mathrm{r}_0 \mathrm{s}_1(\mathbf{V}$[6][c] - $\mathbf{V}$[2][c])\;
      }
      \For{$0 \leq \mathrm{k} \leq N$}{
        t = $\xi_k$\; 
        J$_{00}$ = $\mathbf{E}_0$[j][0] + t$\cdot\mathbf{E}_1$[j][0];\quad J$_{01}$ = $\mathbf{F}_0$[i][0] + t$\cdot\mathbf{F}_1$[i][0]\;
        J$_{10}$ = $\mathbf{E}_0$[j][1] + t$\cdot\mathbf{E}_1$[j][1];\quad J$_{11}$ = $\mathbf{F}_0$[i][1] + t$\cdot\mathbf{F}_1$[i][1]\;
        J$_{20}$ = $\mathbf{E}_0$[j][2] + t$\cdot\mathbf{E}_1$[j][2];\quad J$_{21}$ = $\mathbf{F}_0$[i][2] + t$\cdot\mathbf{F}_1$[i][2]\;
        $\mathbf{J}$ = $\begin{bmatrix}
          \mathrm{J}_{00} & \mathrm{J}_{01} & \mathrm{J}_{02} \\
          \mathrm{J}_{10} & \mathrm{J}_{11} & \mathrm{J}_{12} \\
          \mathrm{J}_{20} & \mathrm{J}_{21} & \mathrm{J}_{22}
        \end{bmatrix}$\tcp*[r]{\color{gray}Get Jacobian Mat. $\mathbf{J}_{ijk}^{(e)} = \frac{1}{8}\mathbf{J}$}
        \KwDCall{\textnormal{\textbf{Jacobi2Geo}}$(\mathbf{J}, \mathrm{i}, \mathrm{j}, \mathrm{k})$}\tcp*[r]{\color{gray}Get Geo. at node (i,j,k)}
      }
  }
  \SetKwProg{Fn}{Function}{}{end}
  \Fn{{\textnormal{\textbf{Jacobi2Geo}}}$(\mathbf{J}, \mathrm{i}, \mathrm{j}, \mathrm{k})$}{
    $\mathrm{K}_{00}$ = $\mathrm{J}_{00}\mathrm{J}_{00}$ + $\mathrm{J}_{10}\mathrm{J}_{10}$ + $\mathrm{J}_{20}\mathrm{J}_{20}$;\quad\quad\ \ $\mathrm{K}_{01}$ = $\mathrm{J}_{00}\mathrm{J}_{01}$ + $\mathrm{J}_{10}\mathrm{J}_{11}$ + $\mathrm{J}_{20}\mathrm{J}_{21}$\;
    $\mathrm{K}_{02}$ = $\mathrm{J}_{00}\mathrm{J}_{02}$ + $\mathrm{J}_{10}\mathrm{J}_{12}$ + $\mathrm{J}_{20}\mathrm{J}_{22}$;\quad\quad\ \ $\mathrm{K}_{11}$ = $\mathrm{J}_{01}\mathrm{J}_{01}$ + $\mathrm{J}_{11}\mathrm{J}_{11}$ + $\mathrm{J}_{21}\mathrm{J}_{21}$\;
    $\mathrm{K}_{12}$ = $\mathrm{J}_{01}\mathrm{J}_{02}$ + $\mathrm{J}_{11}\mathrm{J}_{12}$ + $\mathrm{J}_{21}\mathrm{J}_{22}$;\quad\quad\ \ $\mathrm{K}_{22}$ = $\mathrm{J}_{02}\mathrm{J}_{02}$ + $\mathrm{J}_{12}\mathrm{J}_{12}$ + $\mathrm{J}_{22}\mathrm{J}_{22}$\;
    $\mathrm{J}_\mathrm{det}$ = $\mathrm{J}_{00}(\mathrm{J}_{11}\mathrm{J}_{22}$ - $\mathrm{J}_{21}\mathrm{J}_{12})$ - $\mathrm{J}_{10}(\mathrm{J}_{01}\mathrm{J}_{22}$ - $\mathrm{J}_{21}\mathrm{J}_{02})$ + $\mathrm{J}_{20}(\mathrm{J}_{01}\mathrm{J}_{12}$ - $\mathrm{J}_{11}\mathrm{J}_{02})$\;
    $\lambda_\mathrm{geo}$ = $ 0.125 \cdot w_iw_jw_k\ /\ \mathrm{J}_\mathrm{det}$;\quad\quad  $g_\mathrm{wj}$ = $0.015625 \cdot \mathrm{J}_\mathrm{det} \cdot \mathrm{J}_\mathrm{det}$\; 
    $g_{00}$ = $\mathrm{K}_{11}\mathrm{K}_{22}$ - $\mathrm{K}_{12}\mathrm{K}_{12}$;\quad\quad\quad\quad\ \ \ $g_{01}$ = $\mathrm{K}_{02}\mathrm{K}_{12}$ - $\mathrm{K}_{01}\mathrm{K}_{22}$\;
    $g_{02}$ = $\mathrm{K}_{01}\mathrm{K}_{12}$ - $\mathrm{K}_{02}\mathrm{K}_{11}$;\quad\quad\quad\quad\ \ \ $g_{11}$ = $\mathrm{K}_{00}\mathrm{K}_{22}$ - $\mathrm{K}_{02}\mathrm{K}_{02}$\;
    $g_{12}$ = $\mathrm{K}_{01}\mathrm{K}_{02}$ - $\mathrm{K}_{00}\mathrm{K}_{12}$;\quad\quad\quad\quad\ \ \ $g_{22}$ = $\mathrm{K}_{00}\mathrm{K}_{11}$ - $\mathrm{K}_{01}\mathrm{K}_{01}$\;
    \KwDReturn{$\lambda_\mathrm{geo}, g_{00}, g_{01}, g_{02}, g_{11}, g_{12}, g_{22}, g_\mathrm{wj}$}\;
  }
\end{algorithm}

(1) \textbf{Invariant Components} (Lines 12-14): Since the third column of $\mathbf{J}^{\Phi^{(e)}}(r, s, t)$ in Eq.~\eqref{trilinear_Jacobian} depends only on $r$ and $s$, it follows that the third column of $\mathbf{J}_{ijk}^{(e)}$ depends only on $i$ and $j$, i.e.,
\[
\frac{1}{8} \left([(-1,\ 1) \otimes (1 - \xi_j,\ 1 + \xi_j) \otimes (1 - \xi_i,\ 1 + \xi_i)]  \mathbf{V}^{(e)}\right)^\top.
\]
This term can be computed once and reused within the $k$-loop.

(2) \textbf{Common Terms} (Lines 1-9): The transpose of the first column of $\mathbf{J}_{ijk}^{(e)}$ can be expressed as (the second column is analogous)
\begin{equation}
  \label{trilinear_common_terms}
  \begin{matrix}
     & [(1-\xi_k, 1+\xi_k) \otimes (1-\xi_j, 1+\xi_j) \otimes (-1, 1)] {\mathbf{V}^{(e)}} =       \\
     & \quad [(1, 1) \otimes (1-\xi_j, 1+\xi_j) \otimes (-1, 1)] {\mathbf{V}^{(e)}} + \mathrel{\phantom{-\xi_k}}          \\
     & \quad [(-1, 1) \otimes (1-\xi_j, 1+\xi_j) \otimes (-1, 1)] {\mathbf{V}^{(e)}} \cdot \xi_k.
  \end{matrix}
\end{equation}
Since all terms other than $\xi_k$ depend only on $j$, we precompute four intermediate matrices $\mathbf{E}_0$, $\mathbf{E}_1$, $\mathbf{F}_0$, and $\mathbf{F}_1$ (each of size $N_1 \times 3$), thereby eliminating redundant computation:
\begin{equation}
  \begin{matrix}
    &\mathbf{E}_0 = \mathrel{\phantom{-}}[(1, 1) \otimes (\mathbf{1}-\mathbf{\Xi}_{N}, \mathbf{1}+\mathbf{\Xi}_{N}) \otimes (-1, 1)] \mathbf{V}^{(e)}, \\
    &\mathbf{E}_1 = [(-1, 1) \otimes (\mathbf{1}-\mathbf{\Xi}_{N}, \mathbf{1}+\mathbf{\Xi}_{N}) \otimes (-1, 1)] \mathbf{V}^{(e)}, \\
    &\mathbf{F}_0 = \mathrel{\phantom{-}}[(1, 1) \otimes (-1, 1) \otimes (\mathbf{1}-\mathbf{\Xi}_{N}, \mathbf{1}+\mathbf{\Xi}_{N})] \mathbf{V}^{(e)}, \\
    &\mathbf{F}_1 = [(-1, 1) \otimes (-1, 1) \otimes (\mathbf{1}-\mathbf{\Xi}_{N}, \mathbf{1}+\mathbf{\Xi}_{N})] \mathbf{V}^{(e)}. \\
  \end{matrix}
\end{equation}
Here, $\mathbf{1}$ denotes an $N_1$-dimensional column vector of ones. With these common terms, the six entries in the first and second columns of $\mathbf{J}_{ijk}^{(e)}$ can be re-evaluated with only 12 FLOPs (Lines 17-19).

(3) \textbf{Jacobian Matrix to Geometric Factors} (Lines 22-31): After obtaining the Jacobian matrix $\mathbf{J}$ at a node (Line 20), $\mathbf{K} = {\mathbf{J}}^\top \mathbf{J}$ is computed as an intermediate step for Eq.~\eqref{Jacobian_matrix_to_geometric_factors}. Then,
\begin{equation}
  \label{geometric_factors_from_Jacobian_in_tri_algo}
  |\mathbf{J}| \cdot {\mathbf{J}}^{-1} {\mathbf{J}}^\mathrm{-\top} = |\mathbf{J}| \cdot \mathbf{K}^{-1} = |\mathbf{J}| \cdot \frac{\text{adj}(\mathbf{K})}{|\mathbf{K}|} = \frac{\text{adj}(\mathbf{K})}{|\mathbf{J}|},
\end{equation}
where $\text{adj}(\mathbf{K})$ denotes the adjugate matrix of $\mathbf{K}$. The effects of $1/8$, $|\mathbf{J}|$, and the GLL weights are accumulated into $\lambda_{\mathrm{geo}}$, whereas the geometric factors remain unscaled and must be multiplied by $\lambda_{\mathrm{geo}}$ before use. In summary, Algorithm~\ref{algorithm_geometric_factors_trilinear} recomputes the geometric factors for a single trilinear element with 24 memory references and a total cost of $72N_1 + 45N_1^2 + 80N_1^3$ FLOPs.

For the special case of trilinear elements, or parallelepipeds, Algorithm~\ref{algorithm_geometric_factors_parallelepiped} enables geometric factor recomputation at negligible cost, since the Jacobian matrix is constant across all nodes, i.e., $\mathbf{J}_{ijk}^{(e)} \equiv \mathbf{J}^{(e)}$. Thus, in Eq.~\eqref{Jacobian_matrix_to_geometric_factors}, the geometric factors excluding GLL weights can be obtained with just seven memory references:
$
|\mathbf{J}^{(e)}| \cdot {\mathbf{J}^{(e)}}^{-1} {\mathbf{J}^{(e)}}^{-\top} \ \mathrm{and} \ |\mathbf{J}^{(e)}|.
$
Despite the limited use of parallelepiped elements in HOSFEM, Algorithm~\ref{algorithm_geometric_factors_parallelepiped} serves three purposes: it can be applied to specific problems such as square cavity flow, used to evaluate tensor contraction performance, and employed to probe the performance ceiling of HOSFEM on highly regular meshes. We next introduce performance modeling for kernels with on-the-fly recomputation.

\begin{algorithm}
  \fontsize{9}{9}\selectfont
  \caption{\small Geometric factors recomputation (parallelepiped)}
  \label{algorithm_geometric_factors_parallelepiped}
  \SetKwInOut{Input}{\textbf{Input}}
  \SetKwInOut{Output}{\textbf{Output}}
  \SetKwInOut{Init}{\textbf{Init}}
  \SetKw{KwDStep}{\textbf{step}}
  \SetKw{KwDAnd}{\textbf{and}}
  \SetKw{KwDIn}{\textbf{in}}
  \Input{$h_{0}^{(e)}, ..., h_{6}^{(e)}$: seven geometric factors without GLL weights.
  }
  \For{$0 \leq \mathrm{i,j,k} \leq N$}{
    $W = w_i w_j w_k$\;
    $g_{00}$ = $W\cdot h_{0}^{(e)}$; $g_{01}$ = $W\cdot h_{1}^{(e)}$; $g_{02}$ = $W\cdot h_{2}^{(e)}$\;
    $g_{11}$ = $W\cdot h_{3}^{(e)}$; $g_{12}$ = $W\cdot h_{4}^{(e)}$; $g_{22}$ = $W\cdot h_{5}^{(e)}$; $g_\mathrm{wj}$ = $W\cdot h_{6}^{(e)}$\;
  }
\end{algorithm}

\subsection{Time-Based Roofline Model of AxLocal}
\label{Roofline_model_for_axhelm_subsection}
In this study, we formulate the roofline model from a time-based perspective, estimating the performance bound via computation time $\mathtt{T}_{\mathrm{cmp}}$ and memory access time $\mathtt{T}_{\mathrm{mem}}$ rather than directly adopting the operational intensity form. These two formulations are mathematically equivalent, as shown by Eq.~\eqref{classical_roofline_performance_model}. Fundamentally, the roofline model assesses the minimum achievable execution time by taking the larger of $\mathtt{T}_{\mathrm{cmp}}$ and $\mathtt{T}_{\mathrm{mem}}$. We prefer the time-based formulation because the target platforms involve multiple types of compute cores with different peak performance, making it impractical to reduce the model to the operational intensity expression, and because the kernels contain on-the-fly geometric factor computations $\mathtt{F}_{\mathrm{geo}}$ introduced solely to replace memory accesses, which are not part of the effective computational workload of AxLocal. Including $\mathtt{F}_{\mathrm{geo}}$ in the effective workload would overestimate performance, though its cost is still accounted for in $\mathtt{T}_{\mathrm{cmp}}$.

Naturally, when measuring the performance of AxLocal, we define the \emph{effective performance} $\mathtt{P}_{\mathrm{eff}}$ as the throughput with respect to the effective computational workload, and the \emph{total performance} $\mathtt{P}_{\mathrm{tot}}$ as the throughput including the on-the-fly geometric factor computations. Let $\mathtt{T}_{\mathrm{meas}}$ denote the measured execution time, then  
\[
\mathtt{P}_{\mathrm{eff}} = \frac{\mathtt{F}_{\mathrm{ax}}}{\mathtt{T}_{\mathrm{meas}}}, \quad
\mathtt{P}_{\mathrm{tot}} = \frac{\mathtt{F}_{\mathrm{ax}} + \mathtt{F}_{\mathrm{geo}}}{\mathtt{T}_{\mathrm{meas}}}.
\]
Their corresponding roofline bounds are  
\[
\mathtt{R}_{\mathrm{eff}} = \frac{\mathtt{F}_{\mathrm{ax}}}{\max(\mathtt{T}_{\mathrm{cmp}}, \mathtt{T}_{\mathrm{mem}})}, \quad
\mathtt{R}_{\mathrm{tot}} = \frac{\mathtt{F}_{\mathrm{ax}} + \mathtt{F}_{\mathrm{geo}}}{\max(\mathtt{T}_{\mathrm{cmp}}, \mathtt{T}_{\mathrm{mem}})}.
\]
In fact, the two evaluations differ only by a constant factor given by the ratio of the effective to the total computational workload. The problem then reduces to the evaluation of $\mathtt{T}_{\mathrm{cmp}}$ and $\mathtt{T}_{\mathrm{mem}}$. It is important to note that these quantities must be computed separately for each kernel and for each optimization applied. In particular, (1) $\mathtt{T}_{\mathrm{cmp}}$ should include the time for on-the-fly geometric factor computations and, if applicable, account for the participation of multiple types of compute cores in the computation; (2) $\mathtt{T}_{\mathrm{mem}}$ should reflect the memory access cost after subtracting the amount of data traffic saved compared to a full-memory-access implementation.

\section{Implementation and Optimization}
\label{Implementation_and_optimization_section}
\subsection{Implementation Based on 2D Thread Block}
\label{GPGPU_implementation_subsection}
To achieve efficient execution of the AxLocal kernel on GPGPU architectures, we adopt a 2D thread block layout, as shown in Algorithm~\ref{gpgpu_axlocal_implementation}. Taking the scalar-field case as an example, with $\mathbf{X}^{(e)}$ and $\mathbf{Y}^{(e)}$ each having a single column, each thread block handles one element (loop $e$) and launches an $N_1 \times N_1$ thread layer, where each thread $(i,j)$ maps to one iteration of the $(i,j)$ spatial loop. The innermost loop over $k$ is executed sequentially within each thread to complete the $(i,j,k)$ traversal.

\begin{algorithm}
  \fontsize{9}{9}\selectfont
  \caption{Baseline GPGPU framework for AxLocal using 2D thread blocks}
  \label{gpgpu_axlocal_implementation}
  \SetKwInOut{Input}{\textbf{Input}}
  \SetKwInOut{Output}{\textbf{Output}}
  \SetKwInOut{Init}{\textbf{Init}}
  \SetKw{KwDStep}{\textbf{step}}
  \SetKw{KwDAnd}{\textbf{and}}
  \SetKw{KwDIn}{\textbf{in}}
  \Input{$\mathbf{\Lambda}_0^{(e)}, \mathbf{\Lambda}_1^{(e)}, \mathbf{G}_{00}^{(e)}, ..., \mathbf{G}_{22}^{(e)}, \mathbf{G}_\mathrm{wj}^{(e)}, \mathbf{X}^{(e)}, \hat{\mathbf{D}}_N$.
  }
  \Output{$\mathbf{Y}^{(e)}$ (both input and output are in global memory).
  }
  e, i, j = getIDs()\tcp*[r]{\color{gray} block and thread IDs}
  s\_D[j][i] = $\hat{\mathbf{D}}_N$[j][i]\tcp*[r]{\color{gray}’s\_' means shared memory}
  \For{$0 \leq \mathrm{k} \leq N$}{
    x[k] = $\mathbf{X}^{(e)}$[i,j,k]; y[k] = 0\tcp*[r]{\color{gray}fiber $\mathbf{X}^{(e)}[i,j,:]$}
  }
  \textbf{syncthreads}()\;
  \For{$0 \leq \mathrm{k} \leq N$}{
    $g_{00}, g_{01}, g_{02}, g_{11}, g_{12}, g_{22}, g_\mathrm{wj}, \lambda_0, \lambda_1$ = getFactors(i, j, k)\;
    s\_x[j][i] = x[k]\tcp*[r]{\color{gray}frontal slice ${\mathbf{X}^{(e)}[:,:,k]}^\top$}
    \textbf{syncthreads}()\;
    $x_{0} = \sum_\mathrm{n = 0}^{N}$s\_D[i][n]s\_x[j][n]; $x_{1} = \sum_\mathrm{n = 0}^{N}$s\_D[j][n]s\_x[n][i]\;
    $x_{2} = \sum_\mathrm{n = 0}^{N}$s\_D[k][n]x[n]\;
    s\_r[j][i]$=\lambda_0(g_{00}x_{0}+g_{01}x_{1}+g_{02}x_{2})$\;
    s\_s[j][i]$=\lambda_0(g_{01}x_{0}+g_{11}x_{1}+g_{12}x_{2})$; t$=\lambda_0(g_{02}x_{0}+g_{12}x_{1}+g_{22}x_{2})$\;
    \textbf{syncthreads}()\;
    $y_{0} = \sum_\mathrm{n = 0}^{N}$s\_D[n][i]s\_r[j][n]; $y_{1} = \sum_\mathrm{n = 0}^{N}$s\_D[n][j]s\_s[n][i]\;
    \For{$0 \leq \mathrm{n} \leq N$}{
      y[n] += s\_D[k][n]t\tcp*[r]{\color{gray}$y_2$ accumulated in $k$-loop}
    }
    y[k] += $y_{0} + y_{1} + \lambda_1\cdot g_\mathrm{wj}\cdot$ x[k]\;
  }
  \For{$0 \leq \mathrm{k} \leq N$}{
    $\mathbf{Y}^{(e)}$[i,j,k] = y[k]\;
  }
\end{algorithm}

\begin{figure}[htbp]
  \centering
  \includegraphics[width=0.48\linewidth]{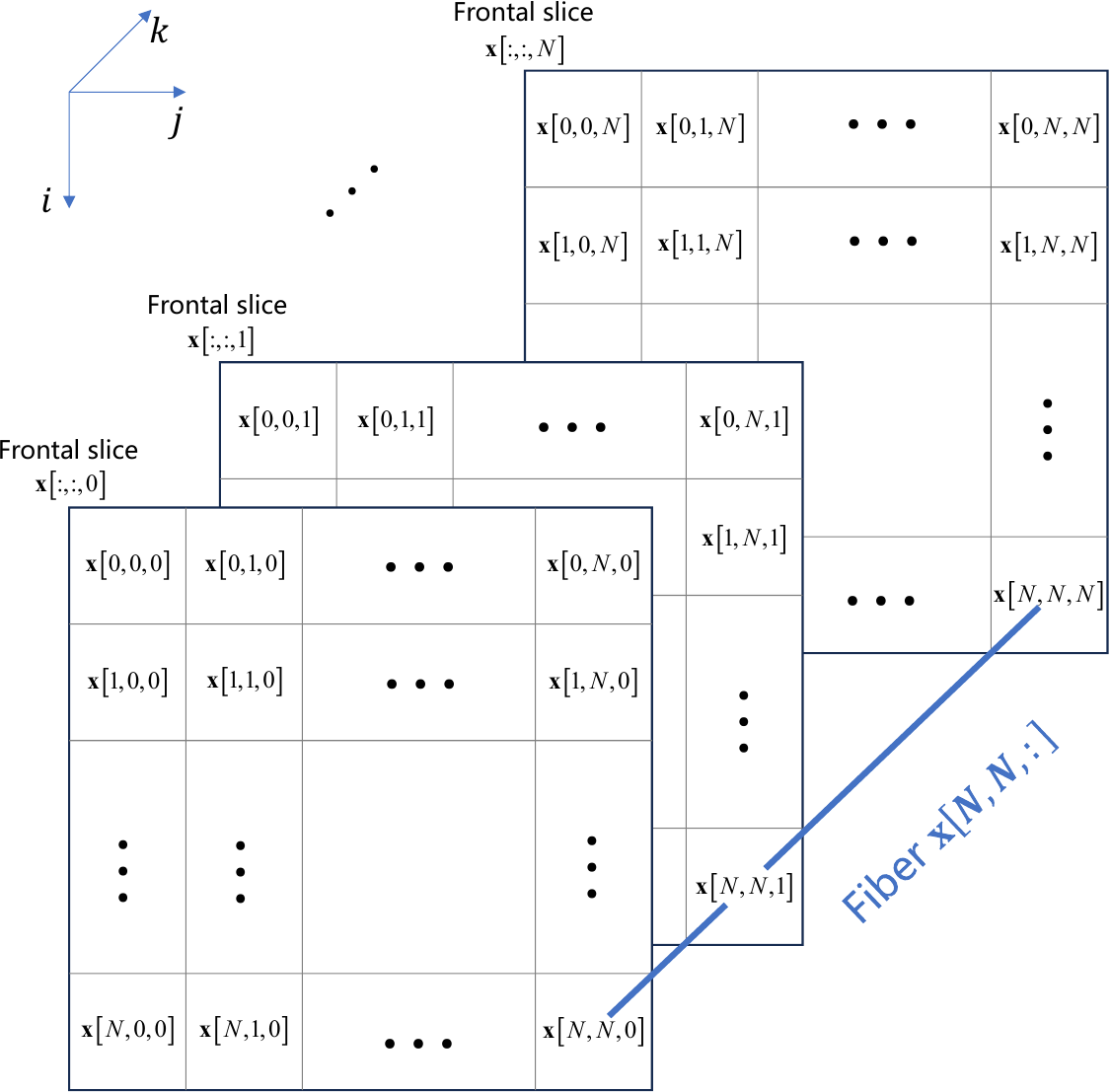}
  \caption{A third-order tensor representation of an $N_1^3$ vector, exhibiting fibers and slices.}
  \label{fig_tensor_structure}
\end{figure}

The input vector $\mathbf{X}^{(e)}$ can be interpreted as a third-order tensor \cite{Tensor_Decompositions_and_Applications}. As illustrated in Fig.~\ref{fig_tensor_structure}, each thread $(i,j)$ owns a fiber $\mathbf{X}^{(e)}[i,j,:]$ along the $k$-dimension. At each $k$-step, a frontal slice $\mathbf{X}^{(e)}[:,:,k]$ is implicitly constructed by gathering the $k$-th element from each fiber, allowing the tensor contractions $\mathbf{D}_r$ and $\mathbf{D}_s$ to proceed layer by layer along the $k$-dimension. This thread-data mapping naturally fits the tensor-product form of AxLocal, enabling fine-grained parallelism and memory access optimization on GPGPUs. Only one $k$-layer resides in shared memory, avoiding the pressure of full 3D tiling. Compared to 3D blocks, the 2D mapping also reduces thread count, improving occupancy and lowering resource contention.

The getFactors interface (Line 7) abstracts the retrieval of geometric and scalar factors. For general elements, these factors are precomputed during the setup stage and stored in global memory; the interface then performs direct memory fetches at each node. For trilinear elements, the $(i,j)$-loop in Algorithm~\ref{algorithm_geometric_factors_trilinear} (Line 10) maps to the $(i,j)$ threads in the implementation, and the $k$-loop (Line 15) is unrolled within getFactors. To avoid redundant operations, both the common terms $\mathbf{E}_0$, $\mathbf{E}_1$, $\mathbf{F}_0$, and $\mathbf{F}_1$ and the invariant components $\mathrm{J}_{02}$, $\mathrm{J}_{12}$, and $\mathrm{J}_{22}$ are computed outside the $k$-loop: the former are stored in shared memory, while the latter are kept in registers.

For vector fields, where both $\mathbf{X}^{(e)}$ and $\mathbf{Y}^{(e)}$ have three columns (i.e., $n_{\mathrm{col}} = 3$), the implementation is entirely analogous. To enable reuse of geometric and scalar factors across different components of the field, the loops over $(i,j,k)$ should be placed outside the loop over field components. Specifically, Line 2 and Line 3 in Algorithm~\ref{AxLocal_algorithm} should be swapped.

\subsection{Optimization of Geometric Factors}
\label{Optimization_of_Geometric_Factors}
This subsection presents geometric factor optimizations for trilinear elements in both Helmholtz and Poisson equations.

\subsubsection{Merging Scalar Factors}
\label{Merging_Scalar_Factors_subsec}
In the Helmholtz equation, AxLocal requires each node to access two scalar coefficients: $\lambda_0$ and $\lambda_1$. Notably, Algorithm~\ref{AxLocal_algorithm} multiplies both coefficients with geometric factors, and Algorithm~\ref{algorithm_geometric_factors_trilinear} introduces a scaling coefficient $\lambda_\mathrm{geo}$ that involves a costly floating-point division on GPGPUs \cite{Floating_Point_Division_on_GPU}. To reduce the cost during kernel execution, we propose pre-applying $\lambda_\mathrm{geo}$ and $g_\mathrm{wj}$ to $\lambda_0$ and $\lambda_1$, respectively, during the setup stage. This yields two new factors:
\[
  \mathbf{\Lambda}_2^{(e)} = \mathbf{\Lambda}_\mathrm{geo}^{(e)}\mathbf{\Lambda}_0^{(e)}, \qquad
  \mathbf{\Lambda}_3^{(e)} = \mathbf{G}_\mathrm{wj}^{(e)}\mathbf{\Lambda}_1^{(e)}.
\]
Before the iterative solve, $\mathbf{\Lambda}_2^{(e)}$ and $\mathbf{\Lambda}_3^{(e)}$ are precomputed and used as input scalars in the AxLocal kernel. As a result, Lines 26-27 in Algorithm~\ref{algorithm_geometric_factors_trilinear} are skipped, and no floating-point division is required during kernel execution.

\subsubsection{Partial Recomputation}
\label{Partial_Recalculation_subsec}
For the Poisson equation, the merging scalar factors optimization is not applicable, since $\lambda_0$, $\lambda_1$, and $g_\mathrm{wj}$ are unused in the AxLocal kernel. However, we can still avoid recomputing $\lambda_\mathrm{geo}$ by moving its calculation from Algorithm~\ref{algorithm_geometric_factors_trilinear} to the setup stage. This partial recomputation strategy replaces expensive in-kernel computations, particularly floating-point division, offering a balanced trade-off between reduced arithmetic complexity and slightly increased memory traffic. Table~\ref{table_overhead_recomputing_geometric_factors} summarizes the overhead of our proposed geometric factor acquisition methods alongside several existing approaches. The comparison illustrates the flexibility in choosing different strategies and highlights that our method incurs significantly lower cost.

\begin{table*}[t]
  \caption{Overhead of acquiring geometric factors (P: Poisson, H: Helmholtz).}
  \label{table_overhead_recomputing_geometric_factors}
  \centerline{\tiny{
      \setlength{\tabcolsep}{2pt}
      \begin{tabular}{|c|c|c|c|c||c|c|c|}
        \hline
        \multirow{2}{*}{Approach}                  & General                      & Parallelepiped                                             & \multicolumn{2}{c||}{Trilinear}                                                      & \multicolumn{3}{c|}{Existing Approaches}                                                                                                                                                          \\
        \cline{2-8}
                                                   & Full Access                  & Algorithm~\ref{algorithm_geometric_factors_parallelepiped} & Algorithm~\ref{algorithm_geometric_factors_trilinear}                                & Section~\ref{Optimization_of_Geometric_Factors} & \cite{GPU_accelerated_HOSFEM_all_hex_meshes} & \cite{NekRS_website} & \cite{High_Fidelity_Flow_Solver_on_FPGA} \\
        \hline
        $\mathtt{F}_\mathrm{geo}$               & 0                            & $7N_1^3$ (P), $8N_1^3$ (H)                               & $72N_1 + 45N_1^2 + 80N_1^3$ & $72N_1 + 45N_1^2 + 60N_1^3$                     & $242N_1^3$                                                  & $296N_1^3$                 & $81N_1^3 + 18N_1^4$                                  \\
        $\frac{\mathtt{M}_\mathrm{geo}}{\mathrm{FPSize}}$ & $6N_1^3$ (P), $7N_1^3$ (H) & 6 (P), 7 (H)                                      & 24                                                                                   & 24 (H), $24 + N_1^3$ (P)            & 24                                                          & 24                         & $24 + N_1^3$                                         \\
        \hline
      \end{tabular}}}
\end{table*}

\subsection{Optimization of Tensor Contractions}
\label{tensor_contraction_optimization_subsection}
Tensor contractions constitute the bulk of computations in AxLocal (Section~\ref{Performance_bottleneck_analysis_of_AxLocal_subsection}). With the memory access bottleneck from geometric factors effectively alleviated, optimization efforts must now shift toward accelerating tensor contractions. Based on the layer-wise computation strategy of the 2D thread block design introduced in Section~\ref{GPGPU_implementation_subsection}, the tensor contractions of $\mathbf{D}_r$, $\mathbf{D}_s$, and $\mathbf{D}_t$ for a fixed $k$ can be interpreted as follows:
\begin{itemize}
  \setlength{\itemsep}{0pt}
  \setlength{\parsep}{0pt}
  \setlength{\parskip}{0pt}
  \item Matrix multiplication: $\hat{\mathbf{D}}_N$ and the $k$-th slice $\mathbf{x}[:,:,k]$.
  \item Matrix multiplication: $\hat{\mathbf{D}}_N$ and the transpose of $\mathbf{x}[:,:,k]$.
  \item Inner product: the $k$-th row of $\hat{\mathbf{D}}_N$ and each fiber $\mathbf{x}[i,j,:]$.
\end{itemize}
Here, $\mathbf{x}$ denotes one column of $\mathbf{X}^{(e)}$, which is a vector of size $N_1^3$.

In summary, the tensor contraction in AxLocal consists of numerous small BLAS operations per element, and its optimization is most effective when guided by small-BLAS tuning techniques. Once global memory access has been minimized, especially through geometric factor optimizations, further performance improvement should focus on reducing shared memory access overhead. In the baseline GPGPU implementation, each thread performs approximately $10N_1^2$ shared memory accesses to complete the tensor contraction (Algorithm~\ref{gpgpu_axlocal_implementation}, Lines 10-11 and 15-17). When considering only the tensor contraction, the ratio of shared memory accesses to floating-point operations reaches $10:12$, indicating substantial shared memory pressure. To address this issue, we present an optimized strategy in Algorithm~\ref{algorithm_tensor_cores_acceleration_tensor_contraction}, where the contractions involving $\mathbf{D}_r$, $\mathbf{D}_s$, and $\mathbf{D}_t$ are optimized separately. Their transposed counterparts, $\mathbf{D}_r^\top$, $\mathbf{D}_s^\top$, and $\mathbf{D}_t^\top$, are treated in a similar fashion.

\begin{algorithm}
  \fontsize{9}{9}\selectfont
  \caption{Tensor Core-based tensor contraction optimization for $N+1 = 8$}
  \label{algorithm_tensor_cores_acceleration_tensor_contraction}
  \SetKwInOut{Input}{\textbf{Input}}
  \SetKwInOut{Output}{\textbf{Output}}
  \SetKwInOut{Init}{\textbf{Init}}
  \SetKw{KwDStep}{\textbf{step}}
  \SetKw{KwDAnd}{\textbf{and}}
  \SetKw{KwDIn}{\textbf{in}}
  \Input{$\mathbf{X}^{(e)}$.}
  e, i, j = getIDs()\tcp*[r]{\color{gray} block and thread IDs}
  warp = (j * (N + 1) + i) / 32\tcp*[r]{\color{gray} warp ID = 0 or 1}
  s\_D[j][i] = c\_D[j][i]\tcp*[r]{\color{gray} get $\hat{\mathbf{D}}_N$ from constant copy}
  \textbf{syncthreads}()\tcp*[r]{\color{gray}each frag from one half of s\_D or $\text{s\_D}^\text{T}$}
  \textbf{LOAD}:\ \ f\_Dleft,\ \ f\_Dright,\ \ f\_DTupper,\ \ f\_DTlower\;
  \For{$0 \leq \mathrm{k} \leq N$}{
    x[k] = $\mathbf{X}^{(e)}$[i,j,k];
  }
  \textbf{syncthreads}()\;
  \For{$\mathrm{k = 0}$ \KwTo $N$ \KwDStep $\mathrm{2}$}{
    s\_x[0][j][i] = x[k];\qquad\qquad\ \ \ s\_x[1][j][i] = x[k + 1]\;
    \textbf{syncthreads}()\;
    \textbf{FILL}: f\_Acc = 0\;
    \textbf{LOAD}: f\_B = upper half of s\_x[warp]\;
    \textbf{MMA}: f\_Acc += f\_Dleft * f\_B\;
    \textbf{LOAD}: f\_B = lower half of s\_x[warp]\;
    \textbf{MMA}: f\_Acc += f\_Dright * f\_B\;
    \textbf{STORE}: s\_s[warp] = f\_Acc\;

    \textbf{FILL}: f\_Acc = 0\;
    \textbf{LOAD}: f\_A = left half of s\_x[warp]\;
    \textbf{MMA}: f\_Acc += f\_A * f\_DTupper\;
    \textbf{LOAD}: f\_A = right half of s\_x[warp]\;
    \textbf{MMA}: f\_Acc += f\_A * f\_DTlower\;
    \textbf{STORE}: s\_r[warp] = f\_Acc\;
    \textbf{syncthreads}()\;

    $x_0$[0] = s\_r[0][j][i];\qquad \qquad\ \ $x_0$[1] = s\_r[1][j][i]\;
    $x_1$[0] = s\_s[0][j][i];\qquad \qquad\ \ $x_1$[1] = s\_s[1][j][i]\;
    $x_2$[0] = $\sum_\mathrm{n = 0}^{N}$c\_D[k][n]x[n];\quad$x_2$[1] = $\sum_\mathrm{n = 0}^{N}$c\_D[k+1][n]x[n]\tcp*[r]{\color{gray} constant copy}
  }
\end{algorithm}

\subsubsection{Optimization Related to $\mathbf{D}_t$}
Since the GLL points, weights, and the differentiation matrix $\hat{\mathbf{D}}_N$ are fixed once the polynomial order $N$ is given, their constant memory copies can be prepared ahead of time. For the tensor contraction involving $\mathbf{D}_t$ (Algorithm~\ref{gpgpu_axlocal_implementation}, Lines 11 and 17), which essentially reduces to a vector inner product, the computation is still executed by general-purpose computing units such as CUDA cores or stream processors. Notably, each thread $(i, j)$ accesses $\hat{\mathbf{D}}_N$ in a uniform pattern that depends only on the indices $k$ and $n$, independent of $i$ and $j$. This property allows us to replace shared memory access ($\text{s\_D}$) with reads from a constant memory copy ($\text{c\_D}$). This approach provides two key benefits: first, constant memory is equipped with a dedicated cache, enabling broadcast to all threads when accessing the same location, thereby reducing latency; second, constant memory is not subject to bank conflicts, further improving access efficiency.

\subsubsection{Tensor Core Optimization for $\mathbf{D}_r$ and $\mathbf{D}_s$}
The tensor contractions involving $\mathbf{D}_r$ and $\mathbf{D}_s$ are mathematically equivalent to matrix multiplications, offering a critical opportunity for optimization. In particular, when the spectral order satisfies $N+1 = 8$, which is the default setting in many production-grade simulations and also the most widely adopted configuration in HOSFEM applications, these contractions can be efficiently offloaded to dedicated matrix-multiplication accelerators. This choice is consistent with the native MMA shape for double precision. Leveraging such units, including Tensor Cores \cite{NVIDIA_Tensor_Core_Programmability}, Matrix Cores \cite{On_the_Rise_of_AMD_Matrix_Cores}, and Cube Units \cite{Ascend_a_Scalable_and_Unified_Architecture}, enables high-throughput execution of small matrix multiplications and yields substantial performance gains. To fully exploit this opportunity, we propose a tailored optimization scheme for this commonly used spectral order (Algorithm~\ref{algorithm_tensor_cores_acceleration_tensor_contraction}). Throughout this paper, the term “Tensor Core” is used generically to denote any such matrix-multiplication accelerator.

Algorithm~\ref{algorithm_tensor_cores_acceleration_tensor_contraction} describes the optimization based on the abstract Warp or Wave Matrix Multiply-Accumulate (WMMA), which provides a warp-level or wavefront-level interface for expressing matrix multiplication \cite{NVIDIA_Tensor_Core_Programmability, On_the_Rise_of_AMD_Matrix_Cores}. This abstraction introduces the notion of fragments, which represent submatrices mapped to thread-local registers. Within this framework, three fundamental operations are defined. The \textbf{LOAD} operation transfers a block of matrix data from shared or global memory into a fragment. The \textbf{MMA} operation performs matrix multiplication between two fragments and accumulates the result into a target fragment. Finally, the \textbf{STORE} operation writes the computed result from a fragment back to memory. These operations are cooperatively executed by all threads within a warp or wavefront. Given that an $8 \times 8$ thread block contains two warps, the $k$-loop is unrolled with a step size of 2 (Line 9), allowing each warp to process one $k$-layer of tensor contractions for $\mathbf{D}_r$ and $\mathbf{D}_s$ independently in each iteration (Lines 12-23). It is worth noting that Algorithm~\ref{algorithm_tensor_cores_acceleration_tensor_contraction} transposes the tensor contraction of $\mathbf{D}_r$, such that the results stored in $\text{s\_r}$ are indexed with $j$ as the row index and $i$ as the column index. This access pattern aligns memory layout with warp execution, which helps reduce shared memory bank conflicts.

With constant memory and Tensor Core optimizations applied, each thread in Algorithm~\ref{algorithm_tensor_cores_acceleration_tensor_contraction} (Lines 12-27) performs only $6N_1$ shared memory accesses to complete the tensor contractions of $\mathbf{D}_r$, $\mathbf{D}_s$, and $\mathbf{D}_t$, and to store the results back into registers. This represents a significant reduction from the $5N_1^2$ accesses required in Algorithm~\ref{gpgpu_axlocal_implementation} (Lines 10-11), thereby greatly relieving shared memory pressure.

\section{Experimental Evaluation}
\label{Experiments_section}
\subsection{Experimental Configurations}
The experiments are conducted on two GPGPU platforms, and the key parameters relevant to our performance analysis are summarized in Table~\ref{table_hardware_configuration}. Released in 2024, the Hygon K100 DCU (Deep Computing Unit) is based on a new general-purpose GPGPU architecture and is highly compatible with the CUDA ecosystem, enabling mainstream AI and HPC frameworks (e.g., PyTorch, TensorFlow, PaddlePaddle) to run efficiently without modification. It integrates 120 compute units (CUs) and supports FP64, FP32, FP16, and INT8 computations, delivering 49 TFLOPS for FP32, 100 TFLOPS for FP16, and 200 TOPS for INT8. The device is equipped with 32 GB of GDDR6 memory, features a PCIe 4.0 $\times$16 interface, has a TDP of 300 W, and adopts a dual-slot, full-height, full-length form factor. While offering higher peak FP64 throughput than the A100, the K100 has a lower memory bandwidth, an important factor that will be revisited in later performance discussions.
\begin{table}[!htb]
  \caption{Experimental hardware configuration.}
  \label{table_hardware_configuration}
  \centerline{\scriptsize{
      \renewcommand{\arraystretch}{0.75}
      \begin{tabular}{|cc|c|c|}
        \hline
        \multicolumn{2}{|c|}{\multirow{2}{*}{\textbf{Configuration}}}                            & A100-Server      & K100-Server                                   \\
        \cline{3-4}
        &               & Host + 2 GPUs    & Host + 4 DCUs                                 \\
        \hline
        \multirow{3}{*}{\rotatebox{90}{\textbf{HOST}}}   & Model            & AMD EPYC 7713         & Hygon C86 7391H       \\
                                                         & Cores            & $2 \times 64$         & $1 \times 32$         \\
                                                         & Memory           & 256 GB                & 256 GB                \\

        \hline
        \multirow{7}{*}{\rotatebox{90}{\textbf{DEVICE}}} & Model            & NVIDIA A100 GPU          & Hygon K100 DCU              \\
                                                         & Interface        & PCI-e 4.0 $\times$ 16 & PCI-e 4.0 $\times$ 16 \\
                                                         & Memory           & 40 GB                 & 32 GB                 \\
                                                         & BW (theoretical) & 1555 GB/s             & 768 GB/s              \\
                                                         & BW (measured)    & 1360 GB/s             & 520 GB/s              \\
                                                         & Tensor Core      & Supported             & Not Supported         \\
                                                         & FP64 TFLOPS      & 9.7 / 19.5 (Tensor)   & 24.5                  \\
        \hline
      \end{tabular}}}
\end{table}

Different studies suggest varying ranges of appropriate $N$, such as $7 \leq N \leq 11$ \cite{High_Fidelity_Flow_Solver_on_FPGA} and $N \leq 10$ \cite{NekRS_website}. Although smaller values are also investigated \cite{Scalability_of_high_performance_PDE_solvers}, $N = 7$ is both the default in NekRS and the most widely adopted in practice \cite{NekRS_first_paper, SC23_HOSFEM, Reducing_communication_in_CG_HOSFEM, Neko_first_paper, Nek5000_RS_Performance_on_Advanced_GPU, High_Fidelity_Flow_Solver_on_FPGA, Accelerate_axhelm_in_Nekbone_on_A64FX}. In this study, we select polynomial orders $\{3, 5, 7, 9\}$ to cover a representative range including the most common choice, and to enable effective utilization of Tensor Cores on the NVIDIA A100 for $N = 7$. The performance of the AxLocal kernels is evaluated using batch sizes $E$ from $2^{16}$ to $2^{20}$, within which performance remains stable, and the optimal value is identified accordingly.

We adopt axhelm, the Nek series implementation of AxLocal, as our baseline since it is among the state-of-the-art GPGPU implementations available. In the Nek series, NekRS is a well-known GPU-optimized HOSFEM-based CFD solver for the NS equations, while NekBench is a widely used benchmarking suite for evaluating the performance of under-development HOSFEM kernels and algorithms \cite{NekRS_first_paper, Coral2_Benchmarks_website, HipBone_2023, A64FX_performance}. NekBench supports both standalone kernel benchmarks and full HOSFEM benchmark (Nekbone), making it an ideal choice for our baseline and experimental environment.

\subsection{Acceleration Effect of Geometric Factor Recomputation}
\begin{figure*}[htbp]
  \centering
  \subfloat[A100]{%
    \includegraphics[width=0.506\linewidth]{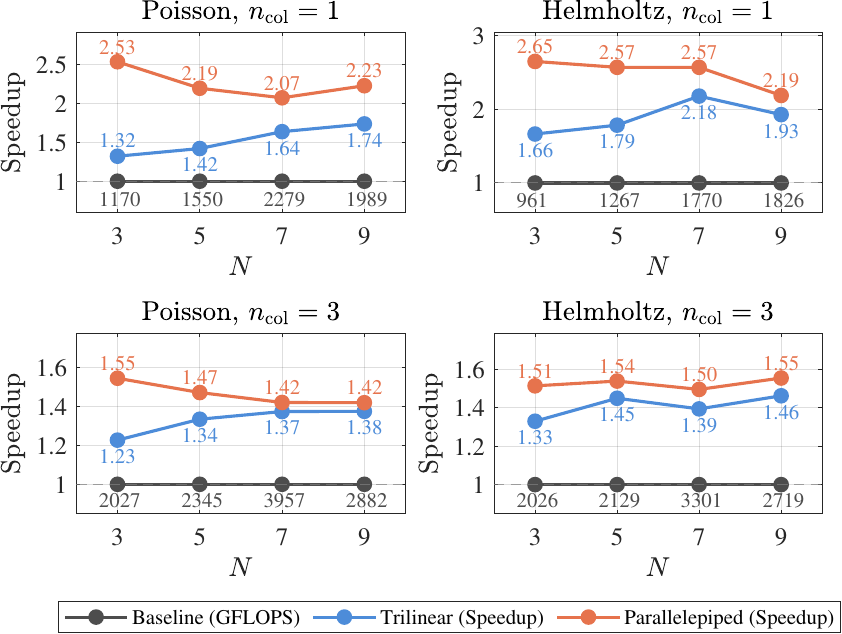}%
  }\hfill
  \subfloat[K100]{%
    \includegraphics[width=0.494\linewidth]{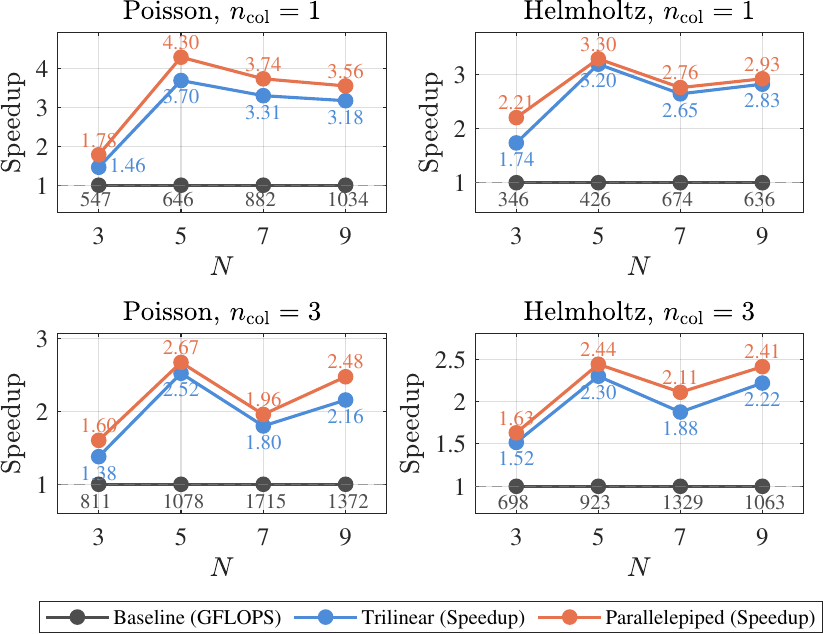}%
  }
  \caption{Speedup of geometric factor recomputation over the baseline on two target platforms.}
  \label{fig_geo_speed_N3579}
\end{figure*}
We first evaluate the performance benefit of on-the-fly recomputing geometric factors for four representative polynomial orders, $N = 3, 5, 7, 9$, on both target platforms. The results, shown in Fig.~\ref{fig_geo_speed_N3579}, report the speedups of the trilinear and parallelepiped element implementations relative to the baseline AxLocal kernel (axhelm). The absolute performance of the baseline is annotated below each data point. In all tested cases, geometric factor recomputation provides significant acceleration over the baseline.

On the K100, both implementations deliver substantial speedup, with the parallelepiped implementation consistently outperforming the trilinear one. The gain is particularly pronounced for $N \geq 5$, where the speedup reaches more than three times over the baseline in some problem configurations. This trend reflects the higher operational intensity and reduced memory traffic achieved by on-the-fly computing geometric factors, which is particularly advantageous for the K100 architecture that, despite having lower memory bandwidth than the A100, offers higher computational capability.

On the A100, similar trends are observed, but the performance gains of both implementations are less pronounced than on the K100, and the performance gap between the parallelepiped and trilinear cases is larger. This is because the A100 has a lower machine balance point, making its performance more sensitive to changes in computational workload. Nevertheless, the performance on the A100 remains higher than that on the K100, owing to its higher memory bandwidth. Since AxLocal is memory-bound, this bandwidth advantage directly translates into higher absolute performance, a point that will be quantitatively analyzed in Section~\ref{Roofline_Analysis_of_AxLocal_subsection}.

Overall, these results confirm that on-the-fly geometric factor recomputation provides a consistent and architecture-agnostic performance benefit. The improvements hold across different polynomial orders and problem types, with larger gains for higher orders and the parallelepiped formulation.

\subsection{Roofline Analysis of AxLocal}
\label{Roofline_Analysis_of_AxLocal_subsection}
\begin{figure}[htbp]
  \centering
  \begin{subfigure}{0.25\textwidth}
    \centering
    \includegraphics[width=\textwidth]{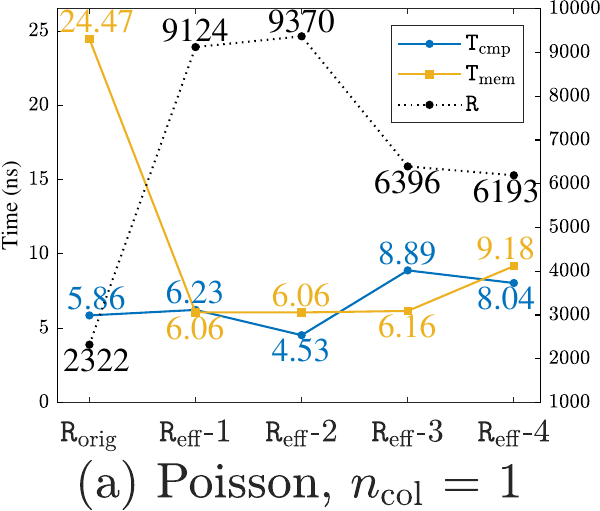}
  \end{subfigure}
  \begin{subfigure}{0.235\textwidth}
    \centering
    \includegraphics[width=\textwidth]{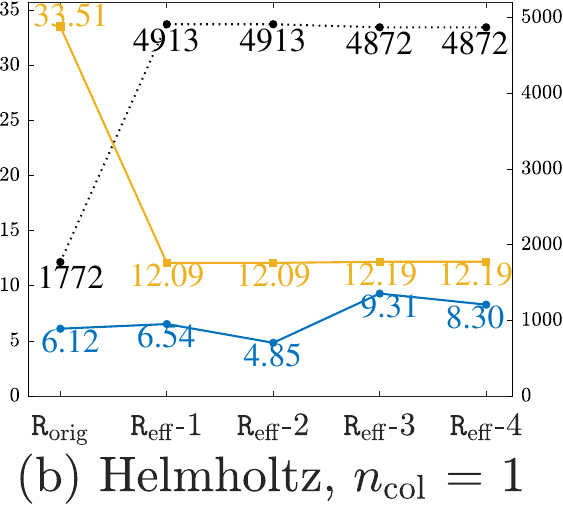}
  \end{subfigure}
  \begin{subfigure}{0.24\textwidth}
    \centering
    \includegraphics[width=\textwidth]{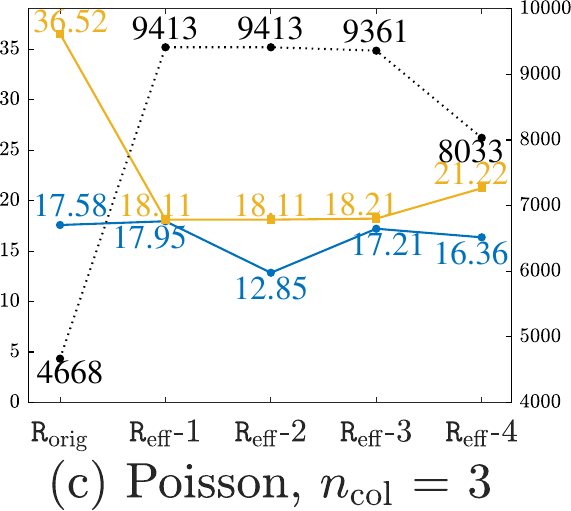}
  \end{subfigure}
  \begin{subfigure}{0.24\textwidth}
    \centering
    \includegraphics[width=\textwidth]{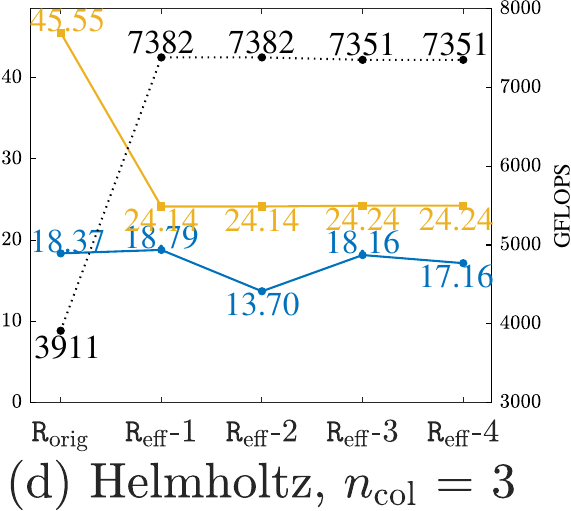}
  \end{subfigure}
  \caption{Roofline analysis of AxLocal on A100 with baseline and optimized kernels.}
  \label{fig_roofline_A100}
\end{figure}

\begin{figure}[htbp]
  \centering
  \begin{subfigure}{0.243\textwidth}
    \centering
    \includegraphics[width=\textwidth]{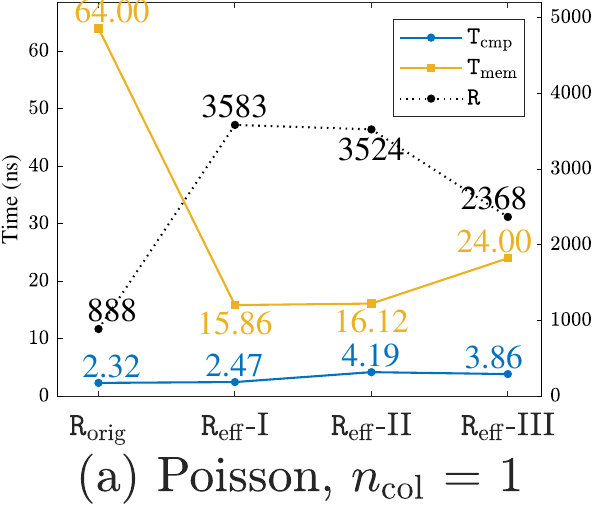}
  \end{subfigure}
  \begin{subfigure}{0.236\textwidth}
    \centering
    \includegraphics[width=\textwidth]{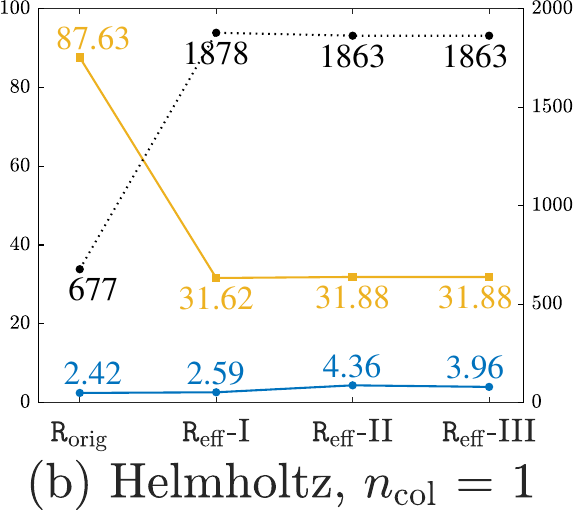}
  \end{subfigure}
  \begin{subfigure}{0.237\textwidth}
    \centering
    \includegraphics[width=\textwidth]{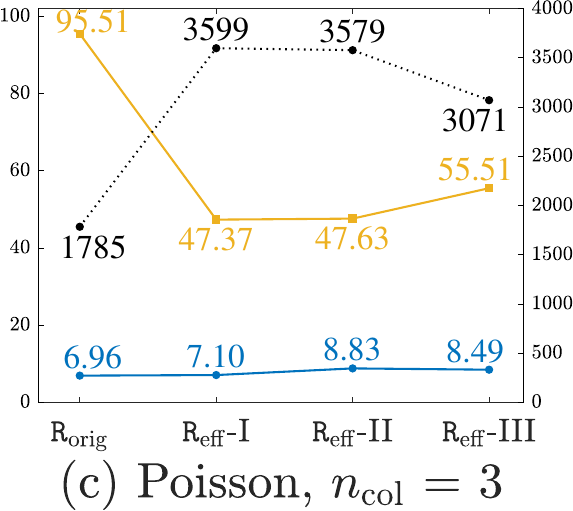}
  \end{subfigure}
  \begin{subfigure}{0.24\textwidth}
    \centering
    \includegraphics[width=\textwidth]{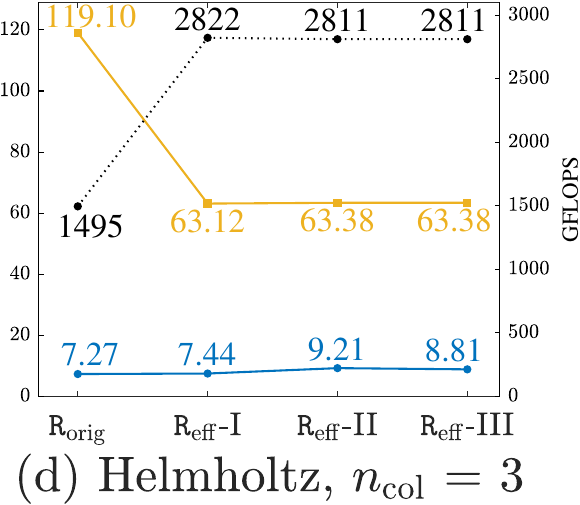}
  \end{subfigure}
  \caption{Roofline analysis of AxLocal on K100 with baseline and optimized kernels.}
  \label{fig_roofline_K100}
\end{figure}

Taking $N=7$ as an example, we compute the roofline bounds using the model in Section~\ref{Roofline_model_for_axhelm_subsection} with the overhead parameters from Table~\ref{table_overhead_recomputing_geometric_factors}. As shown in Fig.~\ref{fig_roofline_A100} and~\ref{fig_roofline_K100}, $\mathtt{R}_\mathrm{orig}$ denotes the baseline kernel. On the A100, $\mathtt{R}_\mathrm{eff}$-1 uses Algorithm~\ref{algorithm_geometric_factors_parallelepiped}; $\mathtt{R}_\mathrm{eff}$-2 adds Tensor Cores to it; $\mathtt{R}_\mathrm{eff}$-3 switches to Algorithm~\ref{algorithm_geometric_factors_trilinear} with Tensor Cores; and $\mathtt{R}_\mathrm{eff}$-4 further applies the optimizations in Section~\ref{Optimization_of_Geometric_Factors}. On the K100, $\mathtt{R}_\mathrm{eff}$-$\mathrm{I}$ uses Algorithm~\ref{algorithm_geometric_factors_parallelepiped}; $\mathtt{R}_\mathrm{eff}$-$\mathrm{II}$ uses Algorithm~\ref{algorithm_geometric_factors_trilinear}; and $\mathtt{R}_\mathrm{eff}$-$\mathrm{III}$ adds the optimizations in Section~\ref{Optimization_of_Geometric_Factors}. We next illustrate the derivation of the effective performance bounds by evaluating the theoretical computation time $\mathtt{T}_{\mathrm{cmp}}$ and memory access time $\mathtt{T}_{\mathrm{mem}}$. Taking the $\mathtt{R}_{\mathrm{eff}}\text{-}3$ configuration on the A100 platform for the Helmholtz equation with $n_\mathrm{col} = 1$ as an example, the memory traffic is $(4N_1^3 + 24)\mathrm{FPSize}$, and $\mathtt{T}_{\mathrm{mem}}$ is obtained by dividing it by the measured bandwidth. The total computation cost consists of $\mathtt{F}_{\mathrm{ax}} = 12N_1^4 + 20N_1^3$ and $\mathtt{F}_{\mathrm{geo}} = 80N_1^3 + 45N_1^2 + 72N_1$, where $8N_1^4$ in $\mathtt{F}_{\mathrm{ax}}$ is executed on Tensor Cores and the remaining FLOPs are executed on general-purpose cores. By substituting the corresponding theoretical peak performance values, $\mathtt{T}_{\mathrm{cmp}}$ can be obtained, and the effective roofline bound is then calculated as $\mathtt{R}_{\mathrm{eff}} = \frac{\mathtt{F}_{\mathrm{ax}}}{\max(\mathtt{T}_{\mathrm{mem}}, \mathtt{T}_{\mathrm{cmp}})}$. The roofline changes because the optimizations modify both memory traffic and computational workload, altering $\mathtt{T}_{\mathrm{mem}}$ and $\mathtt{T}_{\mathrm{cmp}}$.

From the results, except for the Poisson case with $n_\mathrm{col}=1$ on the A100, $\mathtt{T}_{\mathrm{mem}}$ is always greater than $\mathtt{T}_{\mathrm{cmp}}$, with the dominance of $\mathtt{T}_{\mathrm{mem}}$ being particularly pronounced on the K100. This indicates that, in most cases, performance is limited by memory access. Introducing geometric factor recomputation (Algorithms~\ref{algorithm_geometric_factors_trilinear} and~\ref{algorithm_geometric_factors_parallelepiped}) can significantly reduce $\mathtt{T}_{\mathrm{mem}}$; the additional computation cost of Algorithm~\ref{algorithm_geometric_factors_parallelepiped} is almost negligible, while Algorithm~\ref{algorithm_geometric_factors_trilinear} introduces some computational overhead. Nevertheless, both approaches effectively raise the roofline and expand the subsequent optimization space.

On the A100, Tensor Cores significantly shorten $\mathtt{T}_{\mathrm{cmp}}$. For the Helmholtz equation, the merging scalar factors optimization (\ref{Merging_Scalar_Factors_subsec}) is always effective, as it solely reduces computation time; for the Poisson equation, however, the benefit of partial recomputation (\ref{Partial_Recalculation_subsec}) depends on the specific platform: it yields a certain improvement on the A100 but has limited effect on the K100.

Overall, geometric factor recomputation lifts the roofline, alleviating the original memory-bandwidth-bound bottleneck. However, the performance improvement achieved by geometric factor recomputation alone is far less than the increase in the roofline, indicating that there remains substantial room for further optimization. Subsequent performance gains, therefore, rely mainly on optimizing tensor contractions, which are essentially batched small-BLAS operations whose efficiency is size-sensitive and requires careful tuning for each specific case. Next, using the representative case with $N=7$, we illustrate how the quantified roofline guides optimization.

\subsection{Tensor Contraction Performance under Ideal Geometric Conditions}
\label{Evaluation_of_Tensor_Contractions_subsection}
\begin{figure*}[htbp]
  \centering
  \begin{subfigure}[b]{0.72\textwidth}
    \centering
    \raisebox{1.65mm}{\includegraphics[width=\textwidth]{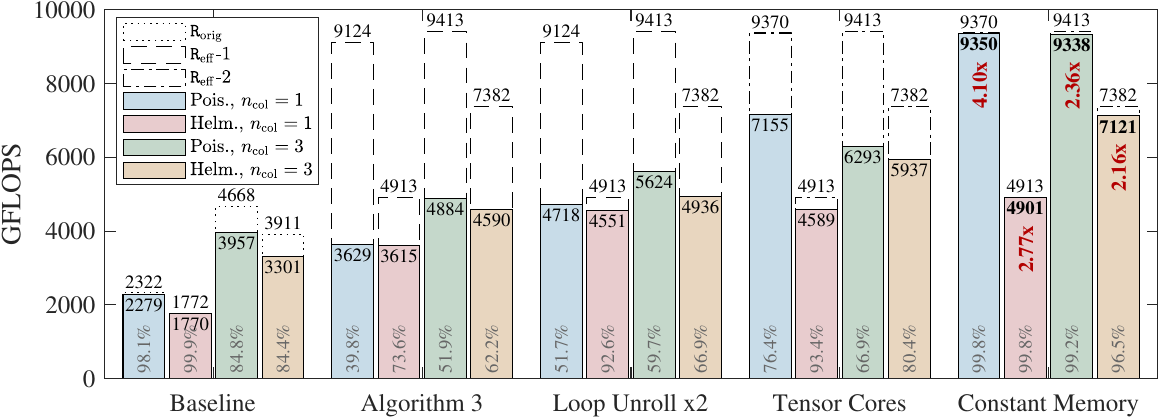}}
    \caption{A100, Parallelepiped}
    \label{fig_axhelm_performance_parallelepiped_A100}
  \end{subfigure}
  \begin{subfigure}[b]{0.27\textwidth}
    \centering
    \includegraphics[width=\textwidth]{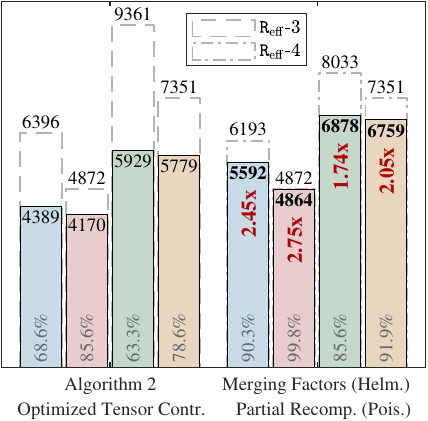}
    \caption{A100, Trilinear}
    \label{fig_axhelm_performance_trilinear_A100}
  \end{subfigure}
  \begin{subfigure}[b]{0.72\textwidth}
    \centering
    \raisebox{1.65mm}{\includegraphics[width=\textwidth]{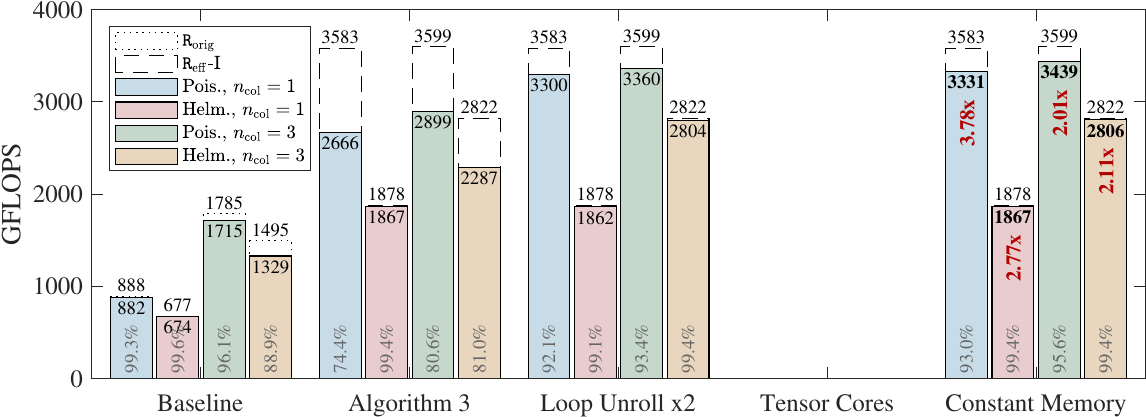}}
    \caption{K100, Parallelepiped}
    \label{fig_axhelm_performance_parallelepiped_K100}
  \end{subfigure}
  \begin{subfigure}[b]{0.27\textwidth}
    \centering
    \includegraphics[width=\textwidth]{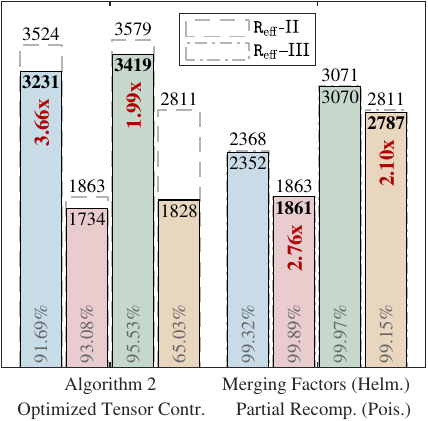}
    \caption{K100, Trilinear}
    \label{fig_axhelm_performance_trilinear_K100}
  \end{subfigure}
  \caption{Performance anatomy of AxLocal kernels (highlighting best performance).}
\end{figure*}

Fig.~\ref{fig_axhelm_performance_parallelepiped_A100} and~\ref{fig_axhelm_performance_parallelepiped_K100} present the effective performance (Section~\ref{Roofline_model_for_axhelm_subsection}) obtained using Algorithm~\ref{algorithm_geometric_factors_parallelepiped} together with successive tensor contraction optimizations. The roofline values shown in these figures are taken from the computation results in Section~\ref{Roofline_Analysis_of_AxLocal_subsection}. The performance of the baseline kernels, after extensive tuning, already reaches 85\%-100\% of $\mathtt{R}_{\mathrm{orig}}$, close to the original roofline. Consequently, even if the tensor contraction computation time were reduced to zero, the overall performance would not improve substantially without modifying the memory access pattern. Although Algorithm~\ref{algorithm_geometric_factors_parallelepiped} elevates the roofline, the actual performance gain is noticeably smaller. This confirms that, while tensor contraction in the baseline has already been well optimized, it still has untapped optimization potential, which becomes visible only once the memory-access bottleneck is alleviated.

Tensor contraction optimizations are applied in sequence: loop unrolling, exploiting Tensor Cores, and leveraging constant-memory copies of the differentiation matrix. As shown by the change from $\mathtt{R}_{\mathrm{eff}}$-1 to $\mathtt{R}_{\mathrm{eff}}$-2 in Fig.~\ref{fig_roofline_A100}, the A100 platform equipped with Tensor Cores still delivers a clear improvement in the measured performance. These optimizations only reduce $\mathtt{T}_{\mathrm{cmp}}$, which is originally smaller than $\mathtt{T}_{\mathrm{mem}}$, and therefore does not change the theoretical performance bound. In particular, the choice of $N + 1 = 8$ is not only an empirical setting from earlier architectures (favoring cache locality and aligning with warp size) but also a configuration that is especially well-suited to newer architectures designed for matrix computation. With these refinements, the roofline efficiency reaches 92.9\%-99.8\%. On the K100, the machine balance point is substantially higher, making computation-side optimization less critical; even without Tensor Cores, simple loop unrolling attains performance close to the elevated roofline. In summary, once the memory bottleneck is mitigated, tensor contraction optimization can reach its performance limit and be directly inherited by AxLocal kernels for trilinear elements.

\subsection{Geometric Factor Optimizations for Trilinear Elements}
\label{Performance_Evaluation_Trilinear}
Fig.~\ref{fig_axhelm_performance_trilinear_A100} and~\ref{fig_axhelm_performance_trilinear_K100} present the performance of AxLocal kernels with Algorithm~\ref{algorithm_geometric_factors_trilinear} and the optimized tensor contraction, along with the evaluation of two optimizations in Section~\ref{Optimization_of_Geometric_Factors}. These figures show a consistent and substantial improvement when Algorithm~\ref{algorithm_geometric_factors_trilinear} is applied in combination with the optimized tensor contraction. Nevertheless, for the A100 platform, all four test cases retain a noticeable gap to the roofline, and on the K100, the Helmholtz case with $n_{\mathrm{col}}=3$ shows similar behavior.

On both platforms, the merging scalar factors optimization raises the performance of all Helmholtz cases to within a few percent of the roofline, as shown in Fig.~\ref{fig_roofline_A100}b and~\ref{fig_roofline_A100}d ($\mathtt{R}_{\mathrm{eff}}\text{-}3 \rightarrow \mathtt{R}_{\mathrm{eff}}\text{-}4$) and Fig.~\ref{fig_roofline_K100}b and~\ref{fig_roofline_K100}d ($\mathtt{R}{\mathrm{eff}}\text{-II} \rightarrow \mathtt{R}{\mathrm{eff}}\text{-III}$). This improvement comes from further reducing computation time while keeping the memory traffic unchanged.

For Poisson cases, the partial recomputation optimization yields performance improvement only on the A100. It also slightly lowers the roofline because of the modest increase in memory traffic. This behavior is clearly reflected in the roofline plots: (1) On the A100 (Fig.~\ref{fig_roofline_A100}a and~\ref{fig_roofline_A100}c: $\mathtt{R}_{\mathrm{eff}}\text{-}3 \rightarrow \mathtt{R}_{\mathrm{eff}}\text{-}4$), partial recomputation effectively rebalances $\mathtt{T}_{\mathrm{cmp}}$ and $\mathtt{T}_{\mathrm{mem}}$ by restoring part of the on-the-fly computation cost to memory fetches. This trade-off is particularly effective on the A100, where computation is relatively slower but memory access is faster, thus shortening the overall execution time. (2) On the K100 (Fig.~\ref{fig_roofline_K100}a and~\ref{fig_roofline_K100}c: $\mathtt{R}_{\mathrm{eff}}\text{-II} \rightarrow \mathtt{R}_{\mathrm{eff}}\text{-III}$), full geometric factor recomputation already maximizes the replacement of memory accesses with computation, yet $\mathtt{T}_{\mathrm{cmp}}$ remains much smaller than $\mathtt{T}_{\mathrm{mem}}$. In this case, partial recomputation would exacerbate the bottleneck and is theoretically ineffective, consistent with the observed performance degradation.

\subsection{Overall Performance Evaluation of HOSFEM Benchmark}
\begin{table}[!htb]
  \caption{HOSFEM Benchmark performance evaluation.}
  \label{table_nekbone_performance_evaluation}
  \centerline{\scriptsize{
      \setlength{\tabcolsep}{2pt}
      \renewcommand{\arraystretch}{0.75}
      \begin{tabular}{|c|c|c|c|c|c|c|c|}
        \hline
                                                               & \multirow{2}{*}{Equation Type}      & \multirow{2}{*}{AxLocal Type} & \multirow{2}{*}{GFLOPS} & \multirow{2}{*}{GDOFS} & \multirow{2}{*}{Accel.} & AxLocal & \multirow{2}{*}{Error \& Iter.} \\
                                                               &                                     &                              &                         &                        &                         & Prop.  &                                 \\
        \hline
        \multirow{12}{*}{\rotatebox{90}{\textbf{A100-Server}, $E$=76$\times$76$\times$76}} & \multirow{3}{*}{Poisson, $n_\mathrm{col} = 1$}   & Baseline                     & 1225                    & 6.51                   & 1x                      & 32\%   & 2.71E-08, 48                    \\
                                                               &                                     & Parallelepiped               & 1594                    & 8.47                   & 1.30x                   & 10\%   & 2.71E-08, 48                    \\
                                                               &                                     & Trilinear                    & 1527                    & 8.11                   & 1.25x                   & 16\%   & 2.71E-08, 48                    \\
        \cline{2-8}
                                                               & \multirow{3}{*}{Helmholtz, $n_\mathrm{col} = 1$} & Baseline                     & 1208                    & 6.20                   & 1x                      & 35\%   & 2.71E-08, 48                    \\
                                                               &                                     & Parallelepiped               & 1560                    & 7.98                   & 1.29x                   & 16\%   & 2.71E-08, 48                    \\
                                                               &                                     & Trilinear                    & 1550                    & 7.93                   & 1.28x                   & 16\%   & 2.71E-08, 48                    \\
        \cline{2-8}
                                                               & \multirow{3}{*}{Poisson, $n_\mathrm{col} = 3$}   & Baseline                     & 1604                    & 8.53                   & 1x                      & 24\%   & 1.42E-08, 49                    \\
                                                               &                                     & Parallelepiped               & 1857                    & 9.87                   & 1.16x                   & 12\%   & 1.43E-08, 49                    \\
                                                               &                                     & Trilinear                    & 1798                    & 9.56                   & 1.12x                   & 16\%   & 1.43E-08, 49                    \\
        \cline{2-8}
                                                               & \multirow{3}{*}{Helmholtz, $n_\mathrm{col} = 3$} & Baseline                     & 1616                    & 8.26                   & 1x                      & 25\%   & 1.42E-08, 49                    \\
                                                               &                                     & Parallelepiped               & 1876                    & 9.60                   & 1.16x                   & 13\%   & 1.43E-08, 49                    \\
                                                               &                                     & Trilinear                    & 1845                    & 9.44                   & 1.14x                   & 14\%   & 1.43E-08, 49                    \\
        \hline
        \hline
        \multirow{12}{*}{\rotatebox{90}{\textbf{K100-Server}, $E$=112$\times$112$\times$56}} & \multirow{3}{*}{Poisson, $n_\mathrm{col} = 1$}   & Baseline                     & 1019                    & 5.42                   & 1x                      & 35\%   & 4.35E-08, 73                    \\
                                                               &                                     & Parallelepiped               & 1362                    & 7.24                   & 1.33x                   & 12\%   & 4.19E-08, 73                    \\
                                                               &                                     & Trilinear                    & 1342                    & 7.14                   & 1.32x                   & 13\%   & 4.19E-08, 73                    \\
        \cline{2-8}
                                                               & \multirow{3}{*}{Helmholtz, $n_\mathrm{col} = 1$} & Baseline                     & 978                     & 5.00                   & 1x                      & 45\%   & 4.34E-08, 73                    \\
                                                               &                                     & Parallelepiped               & 1374                    & 7.03                   & 1.40x                   & 22\%   & 4.19E-08, 73                    \\
                                                               &                                     & Trilinear                    & 1349                    & 6.90                   & 1.38x                   & 22\%   & 4.19E-08, 73                    \\
        \cline{2-8}
                                                               & \multirow{3}{*}{Poisson, $n_\mathrm{col} = 3$}   & Baseline                     & 1274                    & 6.78                   & 1x                      & 31\%   & 2.84E-08, 75                    \\
                                                               &                                     & Parallelepiped               & 1518                    & 8.07                   & 1.19x                   & 17\%   & 2.82E-08, 75                    \\
                                                               &                                     & Trilinear                    & 1508                    & 8.01                   & 1.18x                   & 18\%   & 2.83E-08, 75                    \\
        \cline{2-8}
                                                               & \multirow{3}{*}{Helmholtz, $n_\mathrm{col} = 3$} & Baseline                     & 1271                    & 6.50                   & 1x                      & 33\%   & 2.84E-08, 75                    \\
                                                               &                                     & Parallelepiped               & 1523                    & 7.79                   & 1.20x                   & 18\%   & 2.82E-08, 75                    \\
                                                               &                                     & Trilinear                    & 1502                    & 7.68                   & 1.18x                   & 18\%   & 2.86E-08, 75                    \\
        \hline
      \end{tabular}}}
\end{table}
Table~\ref{table_nekbone_performance_evaluation} summarizes the performance results of Nekbone, the HOSFEM benchmark, obtained with the optimized AxLocal kernels (best-performing configurations) on both target platforms. The proportion of total runtime spent in AxLocal is measured using \texttt{hipprof} and \texttt{nsys}. On the A100 platform, the optimized kernels substantially reduce the AxLocal time overhead, resulting in an overall speedup of 1.25x-1.30x for $n_{\mathrm{col}}=1$ and 1.12x-1.16x for $n_{\mathrm{col}}=3$. On the K100 platform, the corresponding speedups are 1.32x-1.40x and 1.18x-1.20x, respectively. The larger gains for $n_{\mathrm{col}}=1$ on both platforms arise from the higher proportion of runtime spent in AxLocal, with the effect being more pronounced on the K100 due to its larger share. From a numerical perspective, the optimized kernels preserve the accuracy of the original implementation: the error terms remain unchanged to within rounding differences, and the iteration counts are identical across all test cases. This confirms that the optimization strategy maintains solver convergence.

Although the A100-Server has a theoretical FP64 peak of 39~TFLOPS, which is less than half of the K100-Server’s 98~TFLOPS, it achieves higher performance and throughput. This advantage stems from its significantly greater aggregate memory bandwidth (1360~GB/s $\times$ 2 versus 520~GB/s $\times$ 4). Overall, compared with low-order methods, HOSFEM already achieves much higher operational intensity through its matrix-free formulation, while its performance remains primarily constrained by memory bandwidth.

\section{Related Work}
\label{Related_Work_section}
The acceleration of finite element solvers on GPGPUs has been extensively studied \cite{2007_GPU_FEM, 2011_GPU_FEM, 2013_GPU_FEM, 2014_GPU_FEM}. In 2016, Remacle et al. \cite{GPU_accelerated_HOSFEM_all_hex_meshes} proposed the first GPU-accelerated matrix-free HOSFEM algorithm, employing a 3D thread block where each thread processes one node. In 2019, Kasia et al. \cite{Acceleration_tensor_product_operations_in_FEM} introduced a 2D thread block scheme, reducing thread count, supporting larger $N$, and improving shared memory utilization \cite{Optimization_Ax_Nekbone_SC_20_Poster}. This design was later adopted in LibParanumal, NekRS, and Nekbone \cite{Acceleration_tensor_product_operations_in_FEM, Optimization_Ax_Nekbone_SC_20_Poster, NekRS_first_paper}. In 2020, Karp et al. \cite{Optimization_Ax_Nekbone_SC_20_Poster} integrated these tensor-product optimizations \cite{Acceleration_tensor_product_operations_in_FEM} with a hybrid GPU implementation \cite{Nekbone_GPUs_OpenACC_CUDA} into Nekbone. AxLocal has been continuously optimized on different platforms to near-roofline performance \cite{Accelerate_Nekbone_on_FPGA, Accelerate_axhelm_in_Nekbone_on_A64FX, HipBone_2023, Nekbone_GPUs_OpenACC_CUDA, OpenACC_acceleration_Nekbone}.

A well-established optimization strategy in HPC is to replace expensive memory accesses with additional computations when the latter incur lower cost, thereby alleviating memory-bandwidth constraints. Typical examples include gradient checkpointing in machine learning, where intermediate results are recomputed during backpropagation instead of stored \cite{Training_Deep_Nets}. The idea of on-the-fly geometric factor recomputation is a representative instance of this strategy. It was first proposed by Remacle et al. \cite{GPU_accelerated_HOSFEM_all_hex_meshes} in 2016 and later explored in independent implementations \cite{NekRS_website, High_Fidelity_Flow_Solver_on_FPGA}. As summarized in Table~\ref{table_overhead_recomputing_geometric_factors}, existing methods incur high overhead, often resulting in performance degradation or only modest gains over direct access \cite{GPU_accelerated_HOSFEM_all_hex_meshes, Acceleration_tensor_product_operations_in_FEM, Reducing_communication_in_CG_HOSFEM}, and are generally effective only in low-precision settings on platforms where peak compute capability far exceeds memory bandwidth \cite{High_Fidelity_Flow_Solver_on_FPGA}. Consequently, such approaches have been considered impractical \cite{Acceleration_tensor_product_operations_in_FEM} and have received limited attention. Most prior studies instead focus on optimizing tensor contractions, relying on direct memory access for geometric factors while improving loading efficiency via prefetching \cite{HP_SEM_FPGA, Optimization_Ax_Nekbone_SC_20_Poster, Accelerate_axhelm_in_Nekbone_on_A64FX} and packed storage formats \cite{HipBone_2023}.

Building on the 2D thread mapping, we extract the Jacobian matrix’s structure for trilinear elements, reducing recomputation cost via maximal reuse of intermediate results. This enables a significant increase in the achievable roofline, revealing untapped potential in tensor contraction performance. With further optimizations such as merging scalar factors, partial recomputation, Tensor Core acceleration, and constant memory utilization, performance approaches the elevated roofline. These results motivate our work to develop the first practical low-overhead recomputation scheme.

\section{Conclusion}
\label{Conclusion_section}
This work addresses the primary memory-bandwidth bottleneck of the AxLocal kernel in HOSFEM by introducing a practical low-cost geometric factor recomputation scheme for trilinear and parallelepiped elements. This approach is a natural extension of the matrix-free philosophy, which avoids the explicit construction and storage of matrices. While this philosophy is well established in HOSFEM, there remains substantial flexibility in defining the starting point of computation, namely the balance between precomputing intermediate quantities and evaluating them on-the-fly. Most existing implementations repeatedly load geometric factors from memory, reflecting a long-standing belief that these factors are too costly to recompute inside the kernel, leaving a persistent memory-access bottleneck in practical applications. Our approach challenges this assumption by exploiting the tensor-product structure of the trilinear mapping, decomposing the Jacobian into reusable components, and enabling efficient on-the-fly recomputation. We develop a time-based roofline model to quantify the elevated performance bound and to guide further tensor contraction optimizations, including loop unrolling, Tensor Core acceleration, and constant memory utilization. The optimized kernels deliver substantial performance improvements on both A100 and K100 platforms, while preserving numerical accuracy and convergence. These results demonstrate that combining algorithmic reformulation with hardware-aware tuning can remove bottlenecks and fully exploit the performance potential of large-scale high-order simulations.

Future research will focus on developing element-type-adaptive kernels, extending the recomputation approach to other kernels requiring geometric factor access, and integrating the proposed optimizations into practical applications. While multiple polynomial orders have been covered, the focus has been on finely tuning the most commonly used orders; future work will target performance ceilings for other relevant orders. We will also explore low- and mixed-precision implementations and adapt the methods to emerging heterogeneous architectures to better balance computation and memory performance.

\bibliographystyle{elsarticle-num}
\bibliography{Nek-references}

\end{document}